\documentclass{style/nseJournal}
\usepackage{style/defs}

\begin{document}

\title{Collision-Based Hybrid Method for Two-Dimensional Neutron Transport Problems}

\addAuthor{Ben Whewell}{a}
\addAuthor{\correspondingAuthor{Ryan G.~McClarren}}{a}
\correspondingEmail{rmcclarr@nd.edu}

\addAffiliation{a}{Department of Aerospace and Mechanical Engineering \\ University of Notre Dame, Notre Dame, IN 46556, USA}

\addKeyword{Neutron Transport}
\addKeyword{Hybrid Methods}
\addKeyword{Multigroup Approximation}

\titlePage

\begin{abstract}
A collision-based hybrid method for the discrete ordinates approximation of the multigroup neutron transport equation is developed for two-dimensional time-dependent problems. 
At each time step, this algorithm splits the neutron transport equation into two equations, where the external source is part of the uncollided equation and the fission and scattering sources are part of the collided equation.
Low fidelity energy and angular grids are used with the collided transport solution to decrease convergence time while high fidelity grids are used with the uncollided transport solution to limit discretization error.
The hybrid method is shown to be a better solution in terms of both convergence time and accuracy to traditional monolithic coarsening schemes. 
This advantage is demonstrated for two-dimensional time-dependent problems with different materials using a second order temporal discretization scheme.
\end{abstract}

\section{Introduction}
The neutron transport equation (NTE) is used to model how neutrons interact with different target materials \cite{Lewis:1993}. 
This equation calculates the neutron population in a medium over a set period of time by accounting for interactions that add and subtract neutrons to the system. 
One technique for numerically approximating the neutron transport equation is the discrete ordinates (S$_N$) method \cite{Chandrasekhar:2013}.
The S$_N$ method solves for the angular flux at specific angles by using quadrature sets to estimate the integral over the angle.
Time-dependent neutron transport problems are discretized implicitly, where the steady-state problem is solved at each time step before advancing to the subsequent step. 
The steady-state problem is solved using two levels of iteration. 
The inner iteration approximates the scalar flux for each energy group by using a fixed-point source iteration that iterates over all angular directions. 
The self-scattering source is lagged and each update is accomplished using a transport sweep \cite{ODell:1987}.
This sweep requires the inversion of a block triangular operator that accounts for neutron streaming and neutron collisions.
The outer iteration iterate over the energy variable, typically using a Gauss-Seidel method that updates the scattering and fission source terms from the highest to lowest energy levels \cite{Bell:1970}.
Gauss-Seidel is effective in downscattering dominated problems but its convergence time is dependent on the number of neutron interactions that will occur, such as scattering and fission. 
A purely downscattering material without fission will converge in one Gauss-Seidel iteration while materials with upscattering and fission will increase the number of iterations. 
In problems that are optically thick, the number of Gauss-Seidel iterations can become prohibitively large \cite{Lewis:1993}.

Acceleration techniques are used to better handle problems with high number of iterations. 
While accelerators like coarse mesh rebalancing and diffusion synthetic acceleration (DSA), can reduce the cost of the iterative solver, they cannot be used indiscriminately as coarse mesh rebalancing must be concerned with coarse mesh sizes \cite{Lewis:1993} and DSA has difficulty in highly heterogeneous materials \cite{Southworth:2021}.
Other methods that use two-grid, nonlinear diffusion acceleration, and Krylov subspace schemes can improve convergence time with problems containing upscattering and fission \cite{Adams:1993, Anistratov:2013, Slaybaugh:2018} but see a nominal improvement over Gauss-Seidel in downscattering dominant problems \cite{Adams:1993}.
Higher-order time integration approaches have also been explored \cite{Anistratov:2013, Edwards:2011, McClarren:2008, Thoreson:2009}.

Several acceleration techniques can be implemented, but require time to be developed and optimized for their respected transport solvers.
A different approach would be to coarsen the resolution of the neutron flux.
The cost of the multigroup calculation is a function of the energy grid resolution needed, where more energy groups results in a high fidelity solution that comes at the expense of added computational cost.
Smaller energy group structures allow for faster convergence time, although require experts to chose the integration limits to preserve the characteristics of the energy-dependent cross sections \cite{Njoy:2017}.
The angular grid can also be coarsened to improve convergence times. 
Employing a low number of angles comes at the expense of introducing non-physical ray effects that are found in two-dimensional S$_N$ solutions.
These ray effects can be reduced through different techniques \cite{Hauck:2019, Frank:2020} but a common solution is to increase the number of angles \cite{Lewis:1993}, making the angular coarsening approach moot.


Collision-based hybrid algorithms have been a popular area of research that attempt to limit the discretization error for low fidelity solutions \cite{Hauck:2013, Crockatt:2017, Crockatt:2019, Crockatt:2020, Haghighat:2020, Whewell:2023}. 
The transport equation is split into collided and uncollided components as was originally pioneered in the steady-state context \cite{Alcouffe:1977}.
The collided and uncollided components are used with two different energy and angular grids to leverage the fast acceleration times of coarse grids with the accuracy of high resolution grids.
Previous work has shown that the hybrid algorithms are more effective when compared to monolithic discretization schemes. 
This work extends the collision-based hybrid method to the two-dimensional multigroup setting while using a high temporal discretization and a more optimal energy group structure than previous multigroup work \cite{Whewell:2023}.
The hybrid method approach allows for better solutions than its monolithic coarsened solution counterpart while requiring less computational time in high fidelity models.
Faster convergence times over monolithic Gauss-Seidel coarsening scheme are observed when the uncollided and collided equations use the same grid resolution as well. 
This improvement is due to changes in the solver that reduce unnecessary iterations in the inner loop, as explained by a study on over-solving \cite{Senecal:2017}.
These convergence time results are included, however, they have the same error level as the monolithic coarsening scheme.

The organization of this paper is as follows.
Section \ref{sec:neutron-transport} introduces the continuous neutron transport equation, the continuous collided-uncollided split, the multigroup discrete ordinates transport equation, and the collision-based hybrid method.
The memory footprint of the hybrid method, which is larger in two-dimensional problems, will also be discussed in this section.
Section \ref{sec:results} presents two test problems and compares the hybrid method to a monolithic coarsening approach.
Section \ref{sec:conclusion} discusses conclusions and areas of future work.

\section{The Neutron Transport Equation and the Hybrid Formulation} \label{sec:neutron-transport}
The neutron transport equation models the creation and destruction of neutrons as they interact with a surrounding material medium \cite{Lewis:1993}.
In the general equation, there are four components, the time-dependent term, the streaming term, the collision term, and the source term. 
The time-dependent term calculates the change in neutron population over a set time period. 
The streaming term accounts for the neutrons lost to the medium without interacting with the target material, while the collision term accounts for neutrons lost to an interaction with the material. 
The source term can be broken down into scattering, fission, and external sources. 
To fully explain the neutron population, the angular flux $\Psi$ is defined as a function of seven dimensions: space $\bx \equiv (x, y, z) \in D \subset \bbR^3$, angular $\bsOmega \equiv (\theta, \varphi) \in \bbS^2$, energy $E > 0$, and time $t > 0$.
An angular-independent flux, or the scalar neutron flux $\overline \Psi$, takes the angular flux and integrates over the unit sphere as
\begin{align} \label{eq:unit-sphere}
      \overline \Psi(\bx,E,t) = \frac{1}{4 \pi} \int_{\mathbb{S}^2} d\bsOmega \Psi (\bx,\bsOmega,E,t).
\end{align}
Combining the four components, the neutron transport equation becomes
\begin{align} \begin{split} \label{eq:continuous-transport} 
    \left( \frac{1}{v(E)}\frac{\partial}{\partial t} + \bsOmega \cdot \nabla + \sig{t}(\bx, E) \right) \Psi(\bx,\bsOmega,E,t) &= 
    \int_{0}^{\infty}dE' \; \sig{s} (\bx, E' \rightarrow E) \, \overline \Psi(\bx, E', t) \\
    & \hspace{-75pt} + \chi(\bx, E) \int_{0}^{\infty} dE' \; \nu(x, E') \, \sig{f}(\bx, E') \, \overline \Psi(\bx, E', t) +  Q(\bx,\bsOmega, E, t),
\end{split} \end{align}
with the time dependent, streaming, and collision terms on the left hand side, respectively. 
On the right hand side of Eq.~\eqref{eq:continuous-transport}, represents, in order, the scattering source, fission source, and external source.

Each material is characterized by its cross sections, which is the probability that a neutron will interact with the target nucleus of the material \cite{Prinja:2010}.
The NTE is concerned with the scattering $\sig{s}(\bx, E' \rightarrow E)$, fission $\sig{f}(\bx, E')$, and the total $\sig{t}(\bx, E)$ cross sections.
The scattering cross section is the differential scattering cross section for neutrons scattering from energy $E'$ to $E$.
This work assumes isotropic scattering cross sections ($P = 0$) but it can be easily expanded to anisotropic scattering.
The fission cross section is combined with the average number of neutrons created from a fission event $\nu(\bx, E')$ and the probability density that a neutron produced from fission will have energy $E$, represented as $\chi (\bx, E)$.
The neutron velocity $v(E)$ is dependent only on energy $E$ and not on the material through which it is traveling.

The initial condition for Eq.~\eqref{eq:continuous-transport} is 
\begin{align} \label{eq:continuous-initial}
    \Psi(\bx, \bsOmega, E, t=0) = f( \bx, \bsOmega, E) \qquad \text{for} \quad \bx \in D, \quad \bsOmega \in \bbS^2, \quad E > 0,
\end{align} 
where $f$ is the given condition.
The incoming boundary data is represented as
\begin{align} \label{eq:continuous-bcs}
    \Psi(\bx, \bsOmega, E, t) = b( \bx, \bsOmega, E, t) \qquad \text{for} \quad \bx \in \partial D, \quad  \bn(\bx) \cdot \bsOmega < 0, \quad E>0, \quad t >0,
\end{align} 
where $\bn(\bx)$ the unit outward normal at $\bx \in \partial D$, the boundary of $D$.

\subsection{The Continuous Collided-Uncollided Split}
The collision-based hybrid method attempts to accelerate the computation of a numerical solution for Eq.~\eqref{eq:continuous-transport} while maintaining accuracy relative to a monolithic discretization.
The formulation of the hybrid method can be understood at the continuous level as a splitting method.
The NTE is separated into collided and uncollided equations, with distinct angular fluxes calculated for each component and added together to calculate the total angular flux.
The uncollided equation uses the external source as the only source term, while the collided equation uses both the scattering and fission sources.

The uncollided angular flux $\Psiu$ can be calculated from the uncollided neutron transport equation, 
\begin{align} \begin{split} \label{eq:transport-uncollided}
    \left(\frac{1}{v(E)}\frac{\partial}{\partial t} + \bsOmega \cdot \nabla + \sig{t}(\bx, E) \right) \Psiu(\bx, \bsOmega,E,t) &= \Qu(\bx,\bsOmega, E, t),
\end{split} \end{align} 
where $\Qu$ is the external source.
The initial conditions are represented as 
\begin{align}
    \Psiu(\bx, \bsOmega, E, 0) = f(\bx, \bsOmega, E)  \qquad \text{for} \quad \bx \in D, \quad \bsOmega \in \bbS^2, \quad E > 0,
\end{align}
and the incoming boundary data as 
\begin{align}
    \Psiu(\bx, \bsOmega, E, t) = b( \bx, \bsOmega, E, t) \qquad
    \text{for} \quad \bx \in \partial D, \quad  \bn(\bx) \cdot \bsOmega < 0, \quad E > 0, \quad t > 0.
\end{align}

The collided angular flux $\Psic$ is approximated using the collided neutron transport equation, taking the form
\begin{align} \begin{split}
    \left( \frac{1}{v(E)}\frac{\partial}{\partial t} + \bsOmega \cdot \nabla + \sig{t}(\bx, E) \right) \Psic(\bx,\bsOmega,E,t) 
    &=  \int_{0}^{\infty} dE' \; \sig{s} (\bx, E' \rightarrow E) \, \barPsic(\bx, E', t) \\
    &  \hspace{-75pt} + \chi(\bx, E) \int_{0}^{\infty} dE' \; \nu(\bx, E') \, \sig{f}(\bx, E') \, \barPsic(\bx, E', t) + \Qc(\bx, E, t).
\end{split} \end{align} 
The isotropic collided source term $\Qc$ is created from the scattering and fission source of the uncollided flux
\begin{align} \begin{split}
     \Qc(\bx, E, t) &= \int_{0}^{\infty}dE' \; \sig{s} (\bx, E' \rightarrow E) \, \barPsiu(\bx, E', t) \\
     &+ \chi(\bx, E) \int_{0}^{\infty} dE' \; \nu(\bx, E') \, \sig{f}(\bx, E') \, \barPsiu(\bx, E', t).
\end{split} \end{align}
The initial conditions and boundary data are used to calculate the uncollided flux and included in the formulation of the collided source term.
Therefore, the initial condition for $\Psic$ is 
\begin{align}
    \Psic(\bx, \bsOmega, E, 0) = 0 \qquad \text{for} \quad \bx \in D, \quad \bsOmega \in \bbS^2, \quad E > 0,
\end{align} 
and the incoming boundary data is 
\begin{align}
    \Psic(\bx, \bsOmega, E, t) = 0 \qquad \text{for} \quad \bx \in \partial D, \quad \bn(\bx) \cdot \bsOmega < 0, \quad E > 0, \quad t > 0.
\end{align}

A third equation is formulated for the total angular flux $\Psit$, resulting in
\begin{align}\label{eq:total_angular_flux}
    \left(\frac{1}{v(E)} \frac{\partial}{\partial t} + \bsOmega \cdot \nabla + \sig{t} \right) \Psit(\bx,\bsOmega,E,t) = \Qt(\bx, \bsOmega, E, t),
\end{align} 
in which the total external source $\Qt$ is introduced as 
\begin{align} \begin{split}
    \Qt(\bx, \bsOmega, E, t) &=  \int_{0}^{\infty}dE' \; \sig{s} (\bx, E' \rightarrow E) \, \barPsic(\bx, E', t) \\
    &+ \chi(\bx, E) \int_{0}^{\infty} dE' \; \nu (\bx, E') \, \sig{f}(\bx, E') \, \barPsic(\bx, E', t) + \Qc(\bx, \bsOmega, E, t).
\end{split} \end{align}
The initial condition for calculating the total angular flux is 
\begin{align}
    \Psit(\bx, \bsOmega, E, 0) = f(\bx, \bsOmega, E), \qquad \text{for} \quad \bx \in D, \quad \bsOmega \in \bbS^2, \quad E > 0,
\end{align} 
with the incoming boundary data being
\begin{align}
    \Psit(\bx, \bsOmega, E, t) = b(\bx, \bsOmega, E, t) \qquad \text{for} \quad \bx \in \partial D, \quad \bn(\bx) \cdot \bsOmega < 0, \quad E > 0, \quad t > 0. 
\end{align}

When the uncollided and collided fluxes are calculated on the same angular and energy grids, the inclusion of the total flux calculation is redundant, which assumes $\Psit = \Psiu + \Psic = \Psi$.
However, when different discretization methods and grid structures are used to solve for $\Psiu$ and $\Psic$, the inclusion of the total flux calculation no longer becomes trivial.
Employing different grid structures accelerates the convergence time while maintaining accuracy to the full model solution. 
At each time step with the hybrid method, the uncollided flux is approximated on a high fidelity grid while the collided flux is approximated on a low fidelity grid.
These solutions are combined through a remapping process in which the collided flux is reconstructed on the high fidelity grid according to \cite{Hauck:2013}.
Ray effects \cite{Lewis:1993} can unfortunately be introduced with remapping using the discrete ordinates method in multi-dimensional problems.
The inclusion of the total angular flux, as introduced in \cite{Crockatt:2020}, prevents the inclusion of these artifacts when moving from low to high fidelity grids. 
This corrector step allows for easily combining the scalar and angular terms into the total angular flux term.


\subsection{The Multigroup, Discrete Ordinates Equations}
The neutron transport equation is discretized in the angular direction using the discrete ordinates (S$_N$) method \cite{Chandrasekhar:2013}.
The discrete ordinates method solves for the angular flux at specific angles by using quadrature sets to estimate the integral over all angles. 
In this case, the discrete angles are represented as $\bsOmega_{m}$ with a quadrature weight $w_{m}$ where $m \in \cM := \{1, \cdots, M \}$. 
For any integrable function $u$, defined point-wise on the unit sphere $\bbS^2$, the integration is approximated as 
\begin{align}
    \frac{1}{4 \pi} \int_{\bbS^2} d \, \bsOmega \; u(\bsOmega) \approx \sum_{m=1}^{M} w_{m} \, u (\bsOmega_{m}).
\end{align}
There are a number of different quadrature sets that can be used with two-dimensional problems \cite{Jarrell:2011}.
For these results, the product quadrature set is used with the Gauss-Legendre and Gauss-Chebyshev quadratures. 
In addition, it should be noted that when there are $M$ discrete angles in each dimension, there are $M^2$ number of distinct directions for two-dimensional problems.

The multigroup energy discretization attempts to integrate the energy functions over a specific interval \cite{Lewis:1993}. 
A set of finite energies are used for these integral bounds with $0 = E_0 < E_1 < \cdots < E_G = E_{\rm{max}}$ with energy bin widths of $\Delta E_g = E_{g} - E_{g-1}$.
The energy discretization for the angular flux can be approximated as
\begin{align}
    \psi_{m,g}(\bx, t) \approx \int_{E_{g-1}}^{E_{g}} dE \; \Psi(\bx, \bsOmega_m, E, t),\qquad g \in \cG := \{1, \cdots, G\}
\end{align} 
at energy level $g$ and discrete angle $m$.
For the energy discretization for the cross sections, the approximate weight averages of the continuous energy is used such that
\begin{subequations} \begin{align}
    \sig[g' \rightarrow g]{s}(\bx, t) \approx \frac{\displaystyle\int_{E_{g-1}}^{E_{g}} dE  \displaystyle\int_{E_{g'-1}}^{E_{g'}} dE' \; \sig{s} (\bx, E' \rightarrow E) \overline \Psi (\bx, E', t)}{\displaystyle\int_{E_{g'-1}}^{E_{g'}} dE' \; \overline \Psi(\bx, E', t)}  \\ \qquand
    \chi_{g}(\bx) \approx \frac{\displaystyle\int_{E_{g-1}}^{E_{g}} dE \; \chi(\bx, E) \overline \Psi (\bx, E, t)}{\displaystyle\int_{E_{g-1}}^{E_{g}} dE \; \overline \Psi(\bx, E, t)},
\end{align} \end{subequations}
for the $\sig[g]{t}$, $\sig[g' \rightarrow g]{s}$, $\chi_{g}$, $\nu_{g'}$, and $\sig[g']{f}$ values, where the notation $g' \to g$ is used to demonstrate the scattering from energy group $E_{g'}$ to $E_g$.
The approximation comes from the fact that $\overline \Psi(\bx, E, t)$ is not known a priori and an assumed spectral and angular shape of the solution must be used. 
These multigroup cross sections are typically pre-calculated by nuclear data processing software such as NJOY \cite{Njoy:2017} or Fudge \cite{Fudge:2012} and are assumed to be given.

The discretizations for the angular and energy dimensions can be used to rearrange the multigroup, discrete ordinates transport equation as 
\begin{align} \label{eq:transport-discrete}
    \frac{1}{v_g} \frac{\partial}{\partial t} \psi_{m, g} + \bsOmega_{m} \cdot \nabla \psi_{m, g} + \sig[g]{t}\psi_{m, g} = \sum_{g'=1}^{G} \sig[g' \rightarrow g]{s} \barpsi_{g'} +  \chi_{g} \sum_{g'=1}^{G} \nu_{g'} \sig[g']{f} \barpsi_{g'} + q_{m,g}, 
\end{align}
where 
\begin{align}
    \barpsi_{g} = \sum_{m=1}^{M} w_{m} \, \psi_{m,g}
\end{align}
The neutron velocity $v_{g}^{-1}$ is calculated from the relativistic energy formula from \cite[Figure 3]{Bertozzi:1964} at the midpoint of the energy bin. 
The initial condition for the angular flux $\psi_{m, g}$ is 
\begin{align}
    \psi_{m, g} (\bx,0) = f_{m, g} (\bx) \qquad \text{for} \quad \bx \in D, \quad m \in \cM, \quad g \in \cG.
\end{align} 
and the incoming boundary data is
\begin{align}
    \psi_{m, g}(\bx,t) = b_{m, g} ( \bx, t)  \qquad \text{for} \quad \bn(\bx) \cdot \bsOmega_m < 0, \quad m \in \cM, \quad g \in \cG, \quad t > 0,
\end{align} 
where 
\begin{subequations} \begin{align}
    f_{m, g}(\bx) &= \int_{E_{g-1}}^{E_g} dE \; f(\bx, \bsOmega_m, E) \\
    \qquand b_{m, g} &= ( \bx, t)\int_{E_{g-1}}^{E_g} dE \; b(\bx, \bsOmega_m, E, t).
\end{align} \end{subequations} 

The discrete ordinates multigroup neutron transport equation in Eq.~\eqref{eq:transport-discrete} can then be discretized into spatial ($\bx$) and temporal ($t$) components. 
The diamond difference discretization is used for the spatial dimension (see e.g., \cite[Section 4.3]{Lewis:1993} for two-dimensional problems).
Since the diamond differencing scheme for two-dimensional problems is fairly standard with the NTE, the details are omitted. 
It should be noted that a rectangular mesh is used with the diamond difference method and a spatial zone homogenization scheme is used with non-rectangular materials in the medium.
The spatial zone homogenization we use creates a new material based on the percentage of each material in a specific cell. 
The Trapezoidal Rule with Second Order Backward Difference (TR-BDF2) time discretization scheme is used for the temporal discretization.
The TR-BDF2 is a second order method that does not have to deal with increased error from the backward Euler initial time step of the BDF2 method scheme \cite{Nishikawa:2019}.
For each time step $n$, the angular flux at time step $n + \gamma$ is calculated using a Crank-Nicolson temporal discretization before using the flux at the $n$ and $n + \gamma$ time steps with a BDF2 scheme for the $n + 1$ time step \cite{Dharmaraja:2007}.
For this paper, $\gamma = 1/2$ is used \cite{Edwards:2011} and the full implementation is detailed in Algorithm \ref{alg:tr-bdf2} of Appendix \ref{sec:algorithms}.

The discrete approximation of the angular flux $\psi_{m,g}$ is computed using a nested iteration process that includes an outer loop for the energy groups and an inner iteration for the angular directions. 
The procedure is found in Algorithm \ref{alg:multigroup} in Appendix \ref{sec:algorithms}.
A sweeping method \cite{Bell:1970} is the basis of the inner iteration, where the angular flux is known at the cell boundary and used to 
compute the flux exiting the cell\cite{Lewis:1993}.
The inner iteration will sweep in all unique two-dimensional angular directions ($M^2$) using a fixed-point source iteration scheme until convergence of a scalar flux $\barpsi_{g}$ at a specific energy group. 
The convergence of a one group flux is used to update the outer iteration, which uses the Gauss-Seidel method over each energy group.
This method will converge quickly when there is minimal upscattering \cite{Lewis:1993}.

\subsection{The Hybrid Multigroup, Discrete Ordinates Equations} \label{subsec:hybrid}
The multigroup discrete ordinates approximation can be applied to the uncollided, collided, and total flux equations. 
To demonstrate the benefits of the collision-based hybrid method, different angular and energy grid resolutions are used. 
It is assumed that the uncollided and total neutron flux equations employ the same high resolution using $G$ energy groups and $M$ discrete ordinates.
For the collided equation, a coarse grid resolution is used with $\hG$ energy groups and $\hM$ discrete ordinates where $G \geq \hG$ and $M \geq \hM$.
This hat adornment $(\hat{\cdot})$ is used to differentiate the low fidelity, collided parameters from the high fidelity grids. 
For the angular dimension, the collided equations use $\hw_{\hm}$ and $\hOmega_{\hm}$, respectively, for ${\hm} = 1, \dots, \hM$.
For the coarse energy groups, let
\begin{align}
    0 = \Gamma_{0} < \Gamma_{1} \dots < \Gamma_{\hg} \dots < \Gamma_{\hG} = G
\end{align}
for a set of $\hG + 1$ integers with $\hE_{\hg} =  E_{\Gamma_{\hg}}$.  
Then for each $\hg \in \chG$,
\begin{align}
    \Delta \hE_{\hg} = \hE_{\hg} - \hE_{\hg-1} = E_{\Gamma_{\hg}} - E_{\Gamma_{\hat g-1}} = \sum_{g=\Gamma_{\hg-1}+1}^{ \Gamma_{\hg}}
    E_{g} - E_{g-1} = \sum_{g=\Gamma_{\hg-1}+1}^{ \Gamma_{\hg}} \Delta E_g.
\end{align}

The approximation of the uncollided flux, 
\begin{align}
    \psiu_{m,g} \approx \int_{E_{g-1}}^{E_{g}} dE \; \Psiu(\bx, \bsOmega_m, E, t)
\end{align}
uses the uncollided equation
\begin{align} \label{eq:sn-uncollided}
    \frac{1}{v_g} \frac{\partial}{\partial t} \psiu_{m, g} + \bsOmega_m \cdot \nabla \psiu_{m, g} + \sig[g]{t} \psiu_{m, g} = \qu_{m,g}, 
\end{align} 
where the uncollided source is the external source ($\qu_{m,g} = q_{m,g}$). 
The uncollided equation is solved over a time step $[t^n, t^{n+1})$ with initial condition 
\begin{align}
    \psiu_{m, g} (\bx, t^n) = 
    \begin{cases}
    f_{m, g}(\bx), & \bx \in D, \quad m \in \cM, \quad g \in \cG, \quad t^n = 0 \\
    \psit_{m, g}(\bx, t^n_-), &  \bx \in D,\quad m \in \cM, \quad g \in \cG, \quad t^n > 0 
    \end{cases},
\end{align} 
and incoming boundary data 
\begin{align}
    \psiu_{m, g} = b_{m, g} (\bx, t) \qquad \text{for} \quad \bn(\bx) \cdot \bsOmega_m < 0, \quad m \in \cM, \quad g \in \cG \quand t > 0.
\end{align}
The collided transport equation approximates the collided flux
\begin{align}
    \psic_{\hm, \hg} \approx \int_{E_{\hat g-1}}^{E_{\hg}} dE \; \Psic(\bx, \hat{\bsOmega}_m, E, t)
\end{align}
as
\begin{align} \label{eq:sn-collided}
    \frac{1}{v_{\hg}} \frac{\partial}{\partial t} \psic_{\hm, \hg} + \hat \bsOmega_\hm \cdot \nabla \psic_{\hm, \hg} + \hsig[g]{t} \psic_{\hm, \hg} = 
    \sum_{g'=1}^{\hG} \hsig[\hg' \rightarrow \hg]{s} \barpsic_{\hg'} + \hat \chi_{\hg} \sum_{\hg'=1}^{\hG} \hat \nu_{\hg'} \hsig[\hg' \rightarrow \hg]{f} \barpsic_{\hg'} + \qc_{\hg},
\end{align} 
where 
\begin{subequations} \label{eq:hybrid-coarsen-1} \begin{align} 
    v_{\hg} &= \frac{1}{\Delta \hat{E}_{\hg}} \sum_{g=\Gamma_{\hg-1}}^{\Gamma_{\hg}} v_g, \\
    \barpsic_{\hg'} &= \sum_{\hm=1}^{\hM}  \hw_{\hm} \psic_{\hm,\hg'}, \\
    \hat \chi_{\hg} &= \sum_{g=\Gamma_{\hg}+1}^{\Gamma_{\hg+1}} \chi_{g}, \\
    \quand \hat \nu_{\hg'} &= \sum_{g=\Gamma_{\hg}+1}^{\Gamma_{\hg+1}} \nu_{g};
\end{align} \end{subequations}
and the energy-coarsened cross-sections are given by
\begin{subequations} \label{eq:hybrid-coarsen-2} \begin{align} 
    \hsig[\hg]{t} &= \frac{1}{\Delta \hat{E}_{\hg}} \sum_{g=\Gamma_{\hg-1}+1}^{\Gamma_{\hg}} \Delta E_{g} \sig[g]{t}, \\
    \quand \hsig[\hg]{\ell} &= \frac{1}{\Delta \hE_{\hg}} \sum_{g=\Gamma_{\hg-1}+1}^{\Gamma_{\hg}} \sum_{g'=\Gamma_{\hg-1}+1}^{\Gamma_{\hg}} \Delta E_{g'} \sig[g'\rightarrow g]{\ell}, \quad \ell \in \{s, f\}.
\end{align} \end{subequations}
The isotropic source for the collided equation is formulated as 
\begin{align} \label{eq:sn-collided-source}
    \qc_{\hg} = \sum_{g=\Gamma_{\hg-1}+1}^{\Gamma_{\hg}}  \sum_{g'=1}^{G} \sig[g' \rightarrow g]{s} \barpsiu_{g'} + \sum_{g=\Gamma_{\hg-1}+1}^{\Gamma_{\hg}} \chi_{g} \sum_{g'=1}^{G} \nu_{g'} \sig[g' \rightarrow g]{f} \barpsiu_{g'},
\end{align}
where the uncollided scalar flux is
\begin{align}
    \barpsiu_{g} = \sum_{m=1}^M w_m \psiu_{m,g}.
\end{align}
The initial condition and boundary data for solving Eq.~\eqref{eq:sn-collided} over the time step $[t^n, t^{n+1})$ are 
\begin{subequations} \begin{align}
    \psic_{\hm, \hg} (\bx,t^n) &= 0 \qquad \text{for} \quad \bx \in D, \quad \hm \in \chM := \{1, \cdots, \hM \}, \quad \hg \in \chG, \quad t^n \geq 0, \\
    \quand \psic_{\hm, \hg} &= 0 \qquad \text{for} \quad \bn(\bx) \cdot \hOmega_{\hm} < 0, \quad \hm \in \chM, \quad \hg \in \chG, \quad t > 0.
\end{align} \end{subequations}
The total flux, 
\begin{align}
    \psit_{m,g}(\bx,t) \approx \int_{E_{g-1}}^{E_{g}} dE \; \Psit(\bx, \bsOmega_m, E, t)
\end{align} 
is solved according to 
\begin{align} \label{eq:sn-total}
    \frac{1}{v_g} \frac{\partial}{\partial t} \psit_{m, g} + \bsOmega_m \cdot \nabla \psit_{m, \hg} + \sig[g]{t} \psit_{m, g} = \qt_{m,g}.
\end{align} 
For this, the total source is formulated as
\begin{align} \label{eq:sn-total-source}
    \qt_{m,g} = \frac{\Delta E_g}{\Delta \hE_{\hg}}
    \left[\qc_{\hg} + \sum_{\hg'=1}^{\hG} \sig[\hg' \rightarrow \hg]{s} \barpsic_{\hg'} + \chi_{\hg} \sum_{\hg'=1}^{\hG} \nu_{\hg'} \sig[\hg' \rightarrow \hg]{f} \barpsic_{\hg'} \right]
\end{align}
and the scalar collided flux is 
\begin{align}
    \barpsic_{\hg} = \sum_{\hm=1}^{\hM} \hw_{\hm} \psic_{\hm, \hg}.
\end{align}
The value $\hg$ is a unique value such that $\Gamma_{\hg-1} +1 \leq g \leq \Gamma_{\hg}$ or, equivalently, $(E_{g-1},E_{g}) \subset (\hE_{\hg-1},\hE_{\hg})$.
The total flux calculation in Eq.~\eqref{eq:sn-total} takes the initial condition
\begin{align}
    \psit_{m, g} (\bx, t^n) = 
    \begin{cases}
        f_{m, g}(\bx), & \bx \in D, \quad m \in \cM, \quad g \in \cG, \quad t^n = 0 \\
        \psit_{m, g}(\bx, t^n_-), &  \bx \in D, \quad m \in \cM, \quad g \in \cG, \quad t^n > 0 
    \end{cases},
\end{align} 
and incoming boundary data
\begin{align}
    \psit_{m, g} = b_{m, g}(\bx, t) \qquad \text{for} \quad \bn(\bx) \cdot \bsOmega_m < 0, \quad m \in \cM, \quad g \in \cG \quand t > 0.
\end{align}
over a time step $[t^n,t^{n+1})$. Note that Eq. \eqref{eq:sn-total} requires a single iteration because the scattering and fission sources are known from the collided and uncollided steps.

To summarize the collision-based hybrid method in for the discrete ordinates multigroup neutron transport equation, Eqs.~\eqref{eq:sn-uncollided}, \eqref{eq:sn-collided}, and \eqref{eq:sn-total} are solved in succession for each time step.
To map from the high fidelity grid of the uncollided flux to the low fidelity grid of the collided flux, Eq.~\eqref{eq:sn-collided-source} is used. 
Likewise, when mapping from the low fidelity to the high fidelity to approximate the total angular flux, Eq.~\eqref{eq:sn-total-source} is employed.
The solution of the total flux calculation in Eq.~\eqref{eq:sn-total} at the end of each time step acts as the initial condition needed by Eqs.~\eqref{eq:sn-uncollided} and \eqref{eq:sn-total} for the following time step.
The initial condition for \eqref{eq:sn-collided} is set to zero for each time step.

As with the discrete ordinates NTE in Eq.~\eqref{eq:transport-discrete}, diamond differencing and the TR-BDF2 are used for the spatial and temporal discretizations, respectively, of  Eqs.~\eqref{eq:sn-uncollided}, \eqref{eq:sn-collided}, and \eqref{eq:sn-total}.
This paper focuses on implementing the collision-based hybrid method in two-dimensional problems which employs a rectangular grid with details found in \cite{Lewis:1993}.
For non-rectangular materials, a spatial zone homogenization scheme is used, which creates a new composite material out of the percentages of each material that is in a given spatial cell. 
The same spatial zone homogenization scheme and mesh sizes are used for both the monolithic and hybrid equations.
For the temporal discretization with the TR-BDF2 method, the Crank-Nicolson half-step is used to advance the solution from time step $t^n$ to $t^{n+\gamma}$ \cite{Dharmaraja:2007}.
The BDF2 half-step is used to advance the solution to time step $t^{n+1}$ using the solutions at $t^n$ and $t^{n+\gamma}$.
The modified TR-BDF2 method for the collision-based hybrid method is shown in Algorith, \ref{alg:hy-tr-bdf2} of Appendix \ref{sec:algorithms}.
The same time steps and half time steps are used for both the hybrid and monolithic equations.
There are instances where time-dependent cross sections are used in which the appropriate measures can be taken. 
However, it can be assumed in this paper that the cross sections do not change in time, as shown by the exclusion of the time variable in the cross section functions.

The solution procedure for solving the collided flux is similar to the approach used in the standard NTE case.
For this paper, an outer iteration Gauss-Seidel scheme on the energy groups and an inner fixed-point source iteration on the angular directions are employed \cite{Bell:1970, Lewis:1993}. 
The uncollided and the total flux updates can also be solved in this manner, however, do not require any iterations.
This is because the right-hand side of both the uncollided and total flux equations are fixed.
The complete process for implementing the collision-based hybrid method is shown in Algorithm \ref{alg:hy-multigroup} of Appendix \ref{sec:algorithms}.

\subsection{The Hybrid Method Memory Consumption}
The separation of the neutron transport equation into collided and uncollided equations results in the doubling of the number of parameters. 
Each equation requires the angles and quadrature weights ($\bsOmega$, $w$), cross sections ($\sig{t}$, $\sig{s}$, $\sig{f}$), and the fission rate components ($\chi$, $\nu$). 
For one-dimensional multigroup problems in \cite{Whewell:2023}, this doubling of memory requirements did not limit the application even with problems with high numbers of energy groups. 
When expanding to two dimensions, the memory requirements can become large. 
This is exacerbated when applying spatial zone homogenization schemes for non-rectangular materials on a Cartesian grid, as shown with the double chevron problem in Section \ref{sec:results}. 
Although there are 2 materials in the original problem, there are 148 different compositions needed to estimate the chevrons, and therefore, 148 different material properties needed to estimate the neutron flux. 

When implementing the collision-based hybrid method, the memory storage requirement must be further examined. 
For the two-dimensional results presented in this work, the time-dependent hybrid problems are run on personal computers. 
However, there might be instances where this will not be possible, whether due to problem complexity or including the third spatial dimension.
To limit the amount of memory footprint the hybrid method requires, steps can be taken to not require two separate instances of the simulation's material properties.
One technique would be to modify the fixed-source and Gauss-Seidel iterations shown in Algorithm \ref{alg:hy-multigroup} to perform an on-the-fly energy group coarsening scheme.
This would reduce the memory requirements for the problem, without having to reserve space for the cross sections for both the uncollided and collided problem. 
The issue of using triangular meshes would also limit the requirement for large numbers of composite materials.
Although not implemented in this paper, these methods can be seen as a net benefit in the utilization of the collision-based hybrid method.

\section{Numerical Results} \label{sec:results}
Numerical results for the one-dimensional collision-based hybrid method implementation are shown in \cite{Whewell:2023}.
This paper extends the hybrid method to two-dimensional problems, including a higher order discretization technique and energy group coarsening schemes based off of established energy grids from \cite{Leppanen:2015}. 
Given the longer computational times for higher dimensions \cite{Lewis:1993}, these results demonstrate further wall clock time benefits while remaining more accurate than the original coarsening scheme. 
The first problem is a lattice $G = 87$ energy group problem with an americium-beryllium (AmBe) source at the center.
Given the nature of the problem, ray effects become an important issue and the number of discrete ordinates is explored.
The second problem is a double chevron problem with $G = 87$ energy groups that has a time-dependent source entering from the bottom of the medium at $y = 0$ in the 14.1 MeV region.
It should be noted that the high resolution energy grids and material cross sections were created by Fudge \cite{Fudge:2012}. 

\subsection{Hybrid Method Performance Metrics}
The purpose of the collision-based hybrid method is to reduce the required amount of computational time by coarsening the collided energy and angular grids.
This high and low grid resolution separation of the collided and uncollided equations is utilized to increase the accuracy of low fidelity models when compared to a monolithic coarsening procedure. 
To show both the wall clock speed up and the accuracy of the hybrid method over standard techniques, two solutions are shown for each test problem.

The first flux approximation employs the monolithic coarsening of the NTE, referred to as the ``Multigroup'' solution. 
This is the traditional neutron multigroup method with $G$ groups and $M$ angles and coarsened using Eqs.~\eqref{eq:hybrid-coarsen-1} and \eqref{eq:hybrid-coarsen-2}. 
The TR-BDF2 time discretization scheme is used for advancing the time step, as detailed in Algorithm \ref{alg:tr-bdf2}.
For the convergence of the flux at each time step, Gauss-Seidel is used with a tolerance of $\varepsilon_G = 1 \times 10^{-8}$ for the outer iteration and source iteration with a tolerance of $\varepsilon_M = 1 \times 10^{-12}$ for the inner iteration, as described in Algorithm \ref{alg:multigroup}. 
It can be noted that the accuracy is not necessarily monotonic in the number of groups, as there are instances in which smaller numbers of energy groups can yield better results. 
This is caused by the nonlinearity of the procedure as 
a single group calculation can be exact if the true solution is used to compute the group constants \cite[\S 4.3]{Bell:1970}.
Additionally, there can be cancellation of errors in integrated quantities, resulting in these behaviors observed in some test cases. 

The second flux approximation uses the hybrid method, which uses $G$ groups and $M$ angles for the total and uncollided angular fluxes and $\hG$ groups and $\hM$ angles for the collided angular flux.
The TR-BDF2 time discretization scheme is used for advancing the time step in Algorithm \ref{alg:hy-tr-bdf2}. 
The multigroup approximation at each time step uses the algorithm described in Algorithm \ref{alg:hy-multigroup} with iteration tolerances of $\varepsilon_G = 1 \times 10^{-8}$ and $\varepsilon_M = 1 \times 10^{-12}$ for the outer and inner iterations, respectively.
For the hybrid method, it can be assumed that $G > \hG$ and $M > \hM$.
It has been shown that when $G = \hG$ and $M = \hM$, there is a speed up in the convergence by an over-solving process described in \cite{Senecal:2017}.
The downside to this version of the hybrid method, noted as the splitting method, is that there is no benefit to the accuracy of the solution.
The results for the splitting method follow previous trends \cite{Whewell:2023}.

Three metrics are used to show the benefits of the hybrid method. 
The first metric uses the root mean squared error (RMSE) to condense the accuracy down into a single term. 
The inputs to the RMSE metric are the scalar flux, fission rate density, or scatter rate density.
The rate densities calculate the total fission or scattering rate of a specific spatial cell, where the fission rate density is shown as
\begin{align} \label{eq:fission_rate_density}
    {\rm{FRD}} = \; \sum_{g=1}^{G} \left (\chi_{g} \sum_{g'=1}^{G} \nu_{g'} \sig[g' \rightarrow g]{f} \phi_{g'} \right).
\end{align}
To compare the errors of hybrid and multigroup models, the hybrid error is subtracted from the multigroup error, meaning positive values show the hybrid model is more accurate.

The second metric compares the wall clock times as 
\begin{align} \label{eq:wall_clock_diff}
    \tau' = \frac{\tau_{\rm{mg}} - \tau_{\rm{hy}}}{\tau_{\rm{mg}}}, 
\end{align} 
where $\tau_{\rm{mg}}$ and $\tau_{\rm{hy}}$ are the wall clock times required to run the multigroup and hybrid simulations, respectively. 
This format was used to show that positive wall clock time differences indicate the hybrid model converged faster.
To account for variations in the elapsed time, the wall clock was measured for five separate instances and averaged. 

The final metric combines the wall clock time and accuracy into a single metric, which employs a modified figure of merit (FOM) via
\begin{align} \label{eq:figure_of_merit}
    \FOM = \frac{1}{\varepsilon \, \tau},
\end{align} 
where $\varepsilon$ is the RMSE and $\tau$ is the wall clock time.
This is a metric commonly used in the Monte Carlo community \cite{Variansyah:2022} where $\varepsilon^2$ is used in place of $\varepsilon$ in Eq.~\eqref{eq:figure_of_merit} as it is the variance in a statistical estimate.
As a larger figure of merit indicates a more desirable calculation, the FOM difference subtracts the multigroup FOM from the hybrid FOM.
This follows the pattern where positive differences indicate the hybrid model performs better.

\subsection{AmBe Lattice Problem}
\begin{figure}[!ht]
    \centering
    \subfloat[]{\centering \vspace{0.4cm} \includegraphics[width=0.38\textwidth]{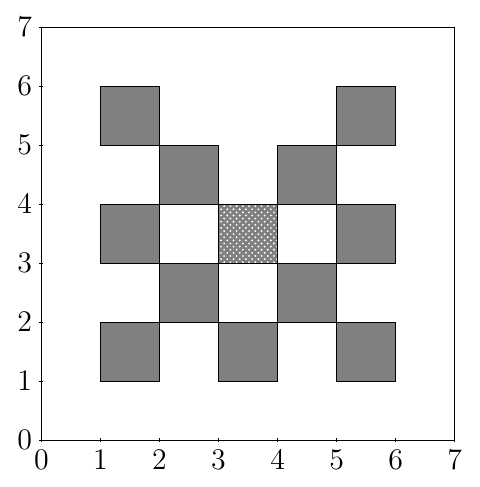} \label{fig:lattice_setup}}
    \subfloat[]{\centering \includegraphics[width=0.58\textwidth]{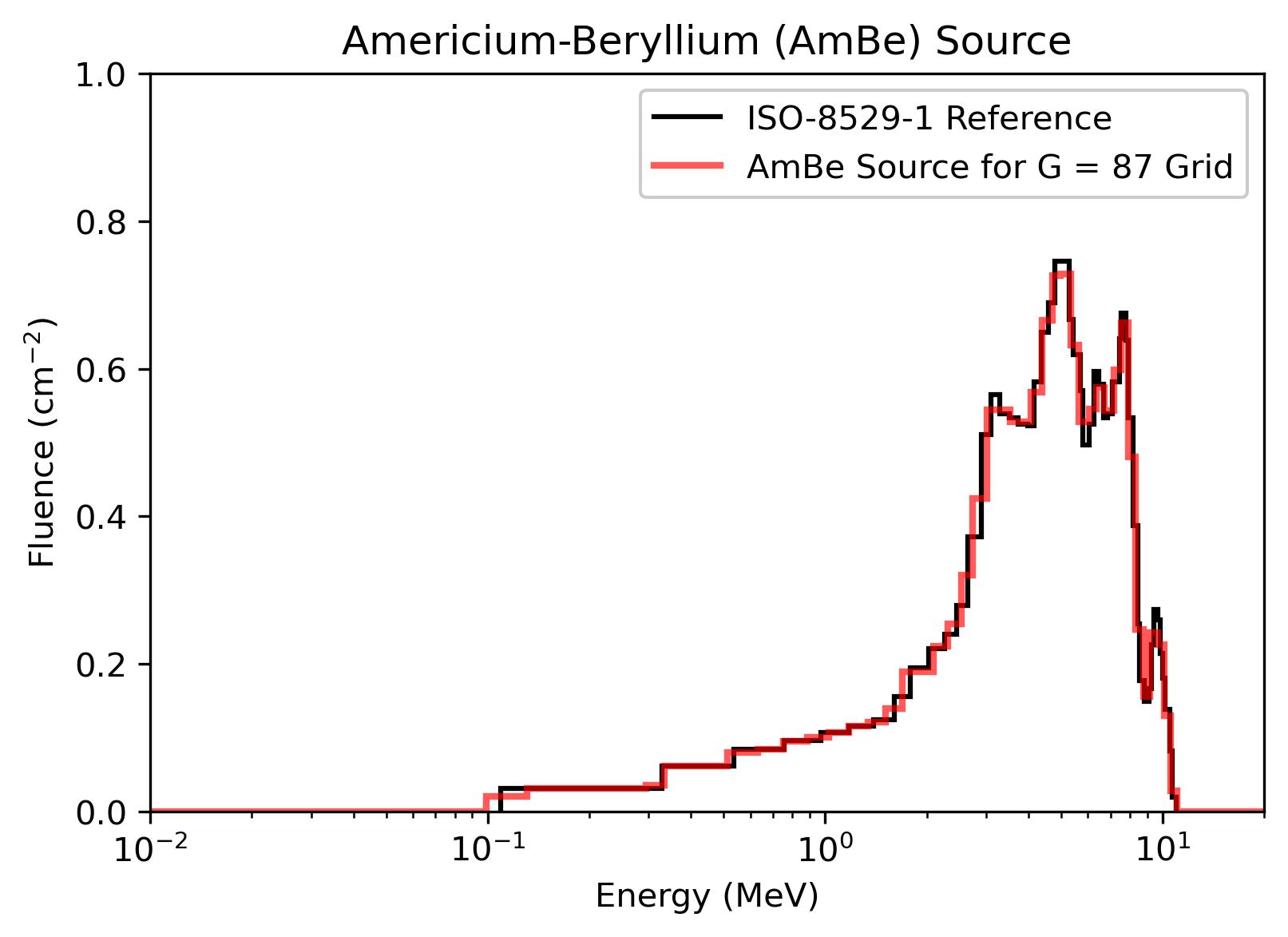} \label{fig:ambe_source}}
    \caption{Setup for the AmBe lattice problem. Figure (a) 100\% enriched $^{235}$U uranium hydride (gray) is surrounded by stainless steel (white) and an AmBe source in the middle of the problem (checkered). Figure (b) is the spectra of the AmBe source in \cite{Mazrou:2010} and fitting it to energy grid used.}
    \label{fig:lattice_problem}
\end{figure}

The first problem is a lattice problem, as shown in Fig.~\ref{fig:lattice_setup}. 
Stainless steel is represented in the white region and 100\% enriched $^{235}$U uranium hydride is the gray region.
There is an americium-beryllium (AmBe) source at the center of the problem, represented by the checkered color and a spectra shown in Fig.~\ref{fig:ambe_source}.
The AmBe source is taken from \cite{Mazrou:2010} and modified to fit the high fidelity $G = 87$ energy grid. 
The medium was $7 \times 7$ cm with vacuum boundaries and the number of spatial cells remained constant where $\Delta x = \Delta y = 0.05$ cm.
Fifty equally spaced time steps were used with a final time of $T = 0.1 \, \mu$s.
The $G = 87$ energy grid and S$_{40}$ angular grid represented the high fidelity reference model.

Ray effects can become a concern for models with low numbers of discrete ordinates, given the construction of this problem. 
There are different techniques to mitigate ray effects, such as \cite{Frank:2020}, but the simplest solution is to increase the number of discrete ordinates \cite{Lewis:1993}.
An S$_{24}$ solution was used for the high fidelity multigroup model, while the high fidelity hybrid models used both S$_{24}$ and S$_{32}$ solutions for the uncollided angles. 
The two hybrid models were used to observe how the number of discrete ordinates affect the accuracy of the problem and the computational impact it has on each model's figure of merit.

\begin{figure}[!ht]
    \centering
    \subfloat[]{\centering \includegraphics[width=0.49\textwidth]{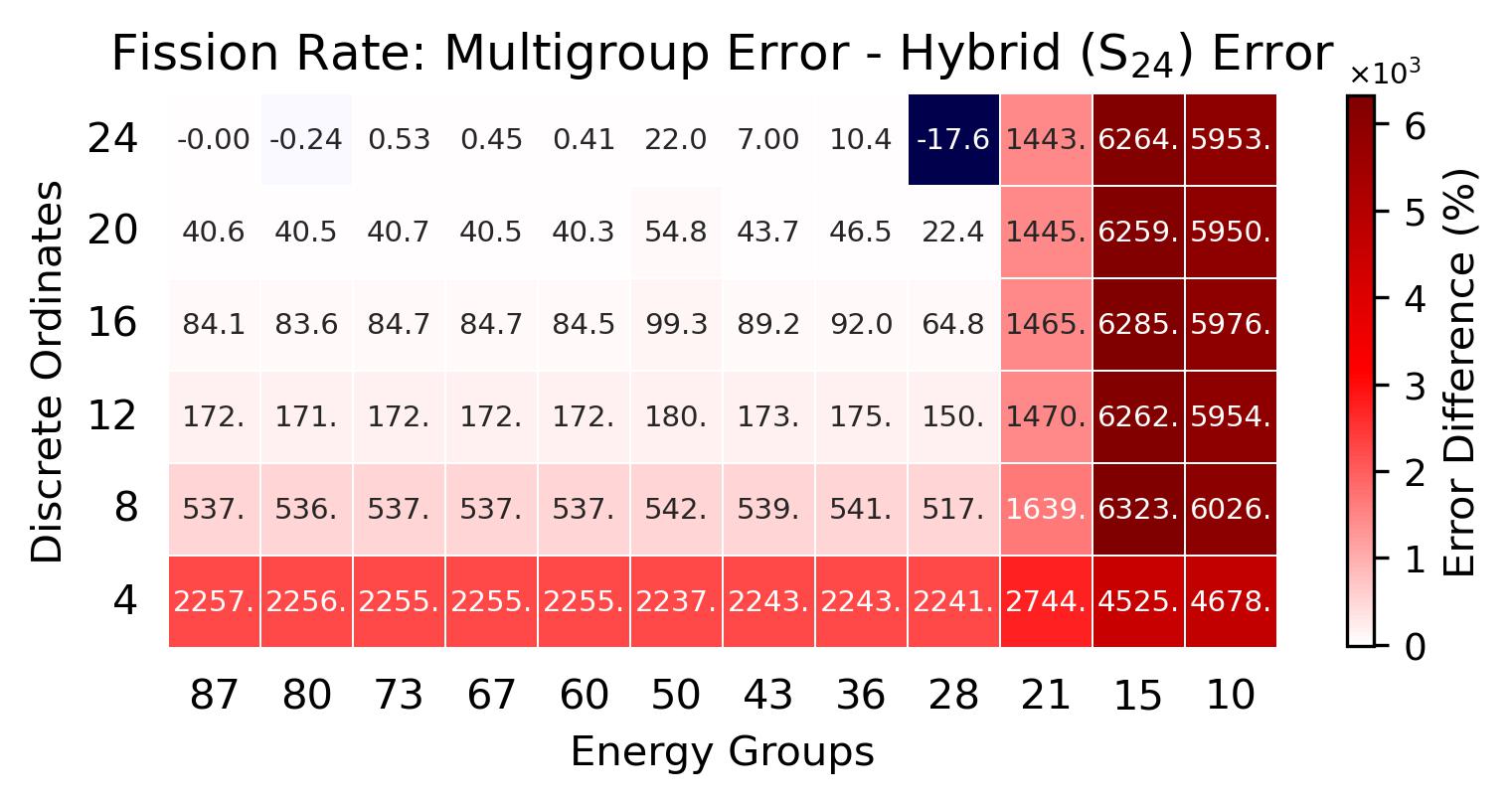} \label{fig:lattice_fission_02}} 
    \subfloat[]{\centering \includegraphics[width=0.49\textwidth]{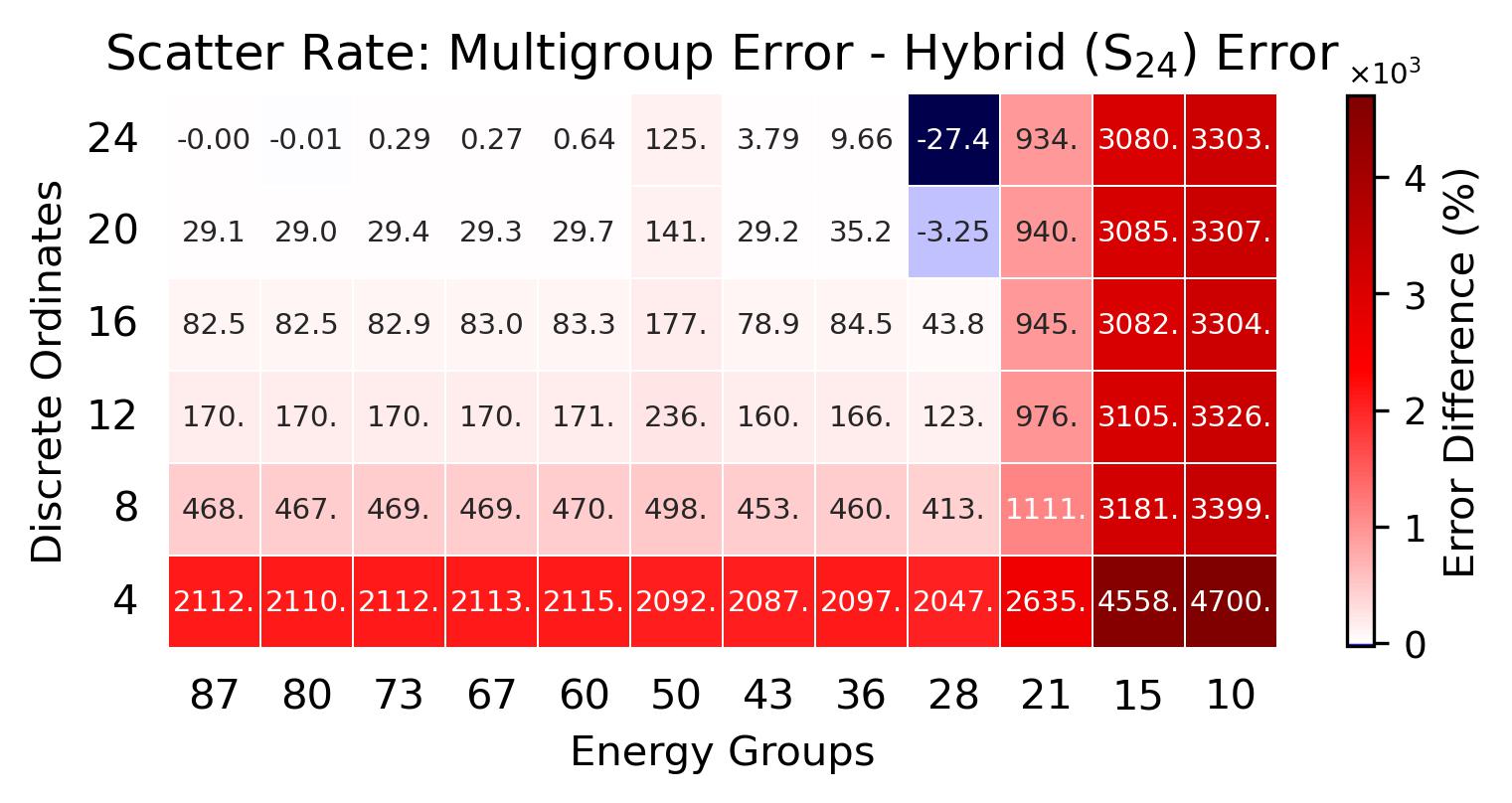} \label{fig:lattice_scatter_02}} 
    \caption{The error difference comparison for the AmBe lattice problem between multigroup models and S$_{24}$ uncollided hybrid models. Figure (a) is the fission rate density RMSE percent difference. Figure (b) is the scattering rate density RMSE percent difference. }
    \label{fig:lattice_error}
\end{figure}

The accuracy comparison between the multigroup and the S$_{24}$ uncollided hybrid models is shown in Fig.~\ref{fig:lattice_error}.
The fission rate density RMSE difference in Fig.~\ref{fig:lattice_fission_02} and the scattering rate density RMSE difference in Fig.~\ref{fig:lattice_scatter_02} show the percent difference in comparison to the S$_{40}$ reference solution.
For both the fission and scattering rate density, the hybrid model is significantly more accurate with low fidelity models in both angle and energy.
The benefit of the hybrid method decreases as higher fidelity models are compared, although models like $G = 87$, $M = 12$ show the hybrid model is still 170\% more accurate than its multigroup counterpart.
The instances where the multigroup model is more accurate, such as $G = 28$, $M = 24$, are most likely caused by the energy group coarsening scheme and error cancellation.

\begin{figure}[!ht]
    \centering
    \subfloat[]{\centering \includegraphics[width=0.49\textwidth]{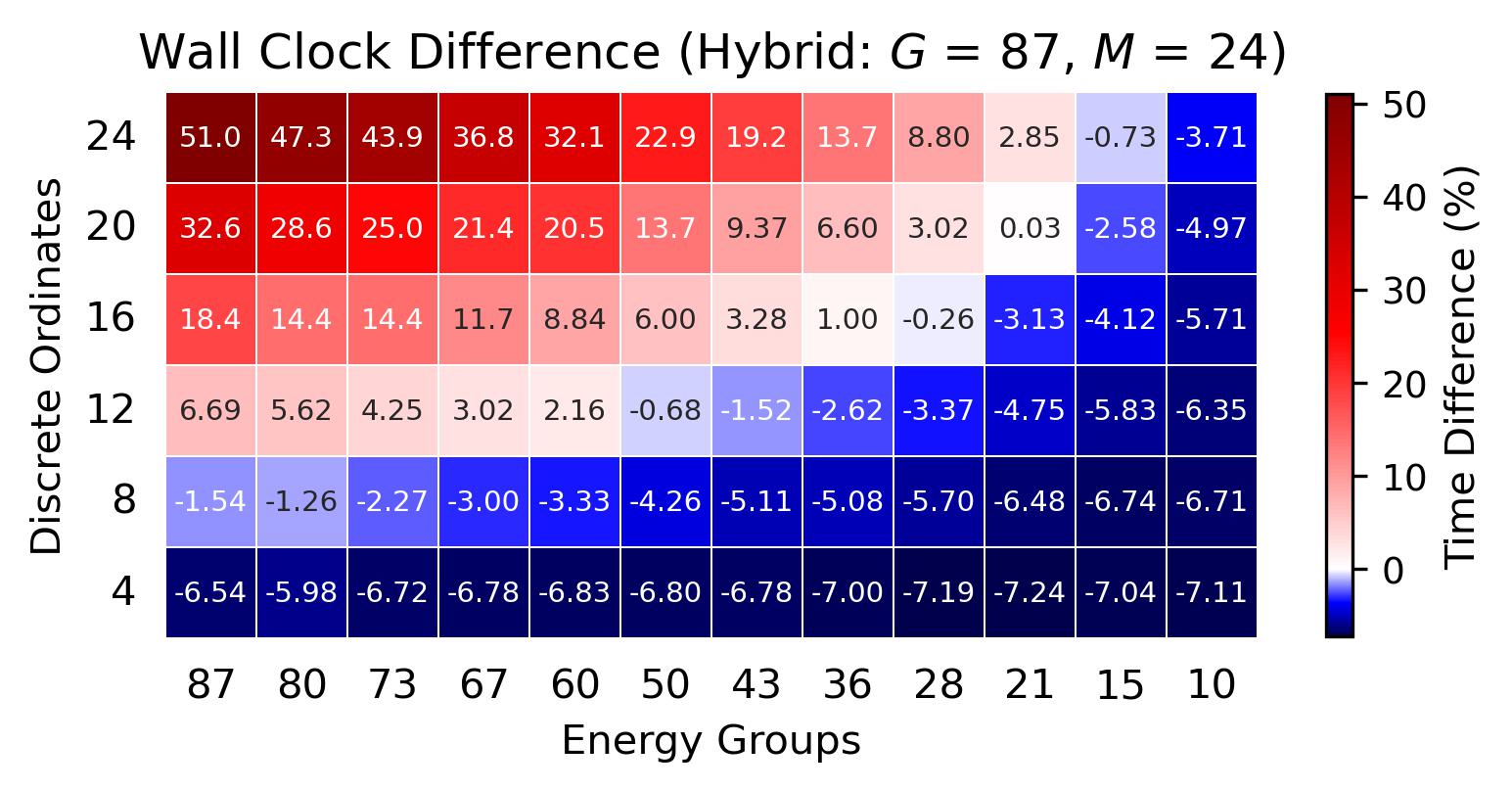} \label{fig:lattice_time_02}}
    \subfloat[]{\centering \includegraphics[width=0.49\textwidth]{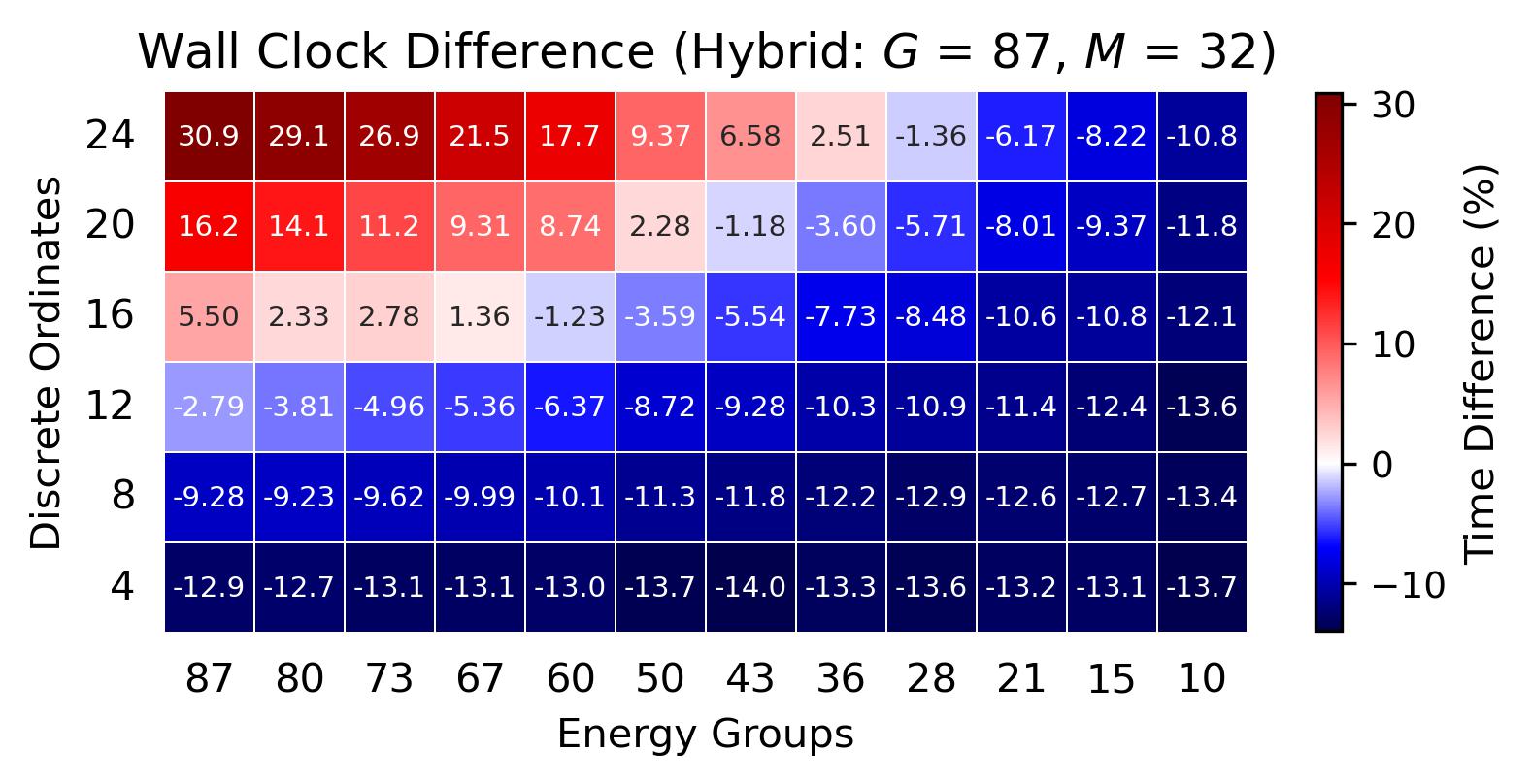} \label{fig:lattice_time_01}} 
    \caption{The wall clock time percent difference comparison for the AmBe lattice problem. Figure (a) uses the S$_{24}$ uncollided hybrid solution. Figure (b) uses the S$_{32}$ uncollided hybrid solution.}
    \label{fig:lattice_time}
\end{figure}

The wall clock time comparison, as described in Eq.~\eqref{eq:wall_clock_diff}, is shown in Fig.~\ref{fig:lattice_time}.
Fig.~\ref{fig:lattice_time_02} uses the S$_{24}$ uncollided angular grid for the hybrid model. 
For these comparisons, the hybrid method converges faster than the multigroup method when utilizing high fidelity angular and energy grids. 
This wall clock benefit disappears for the low fidelity models, most likely due to the uncollided and corrector steps described in Algorithm \ref{alg:hy-multigroup}.
While the convergence is dependent on the collided equation, the uncollided equation is still required to perform sweeps over all angles and energy groups.
This means that there is an additional $G \times M^{2}$ loop performed twice at each time step, once for the uncollided equation and a second for the corrector step.
While it does not affect the wall clock times of the high-fidelity models, it becomes apparent with the low-fidelity models.
Fig.~\ref{fig:lattice_time_01}, which uses the S$_{32}$ uncollided hybrid model, shows similar trends, albeit with an overall decrease in wall clock efficiency. 
This can be expected as the number of uncollided and corrector step iterations required is almost doubled.

\begin{figure}[!ht]
    \centering
    \subfloat[]{\centering \includegraphics[width=0.49\textwidth]{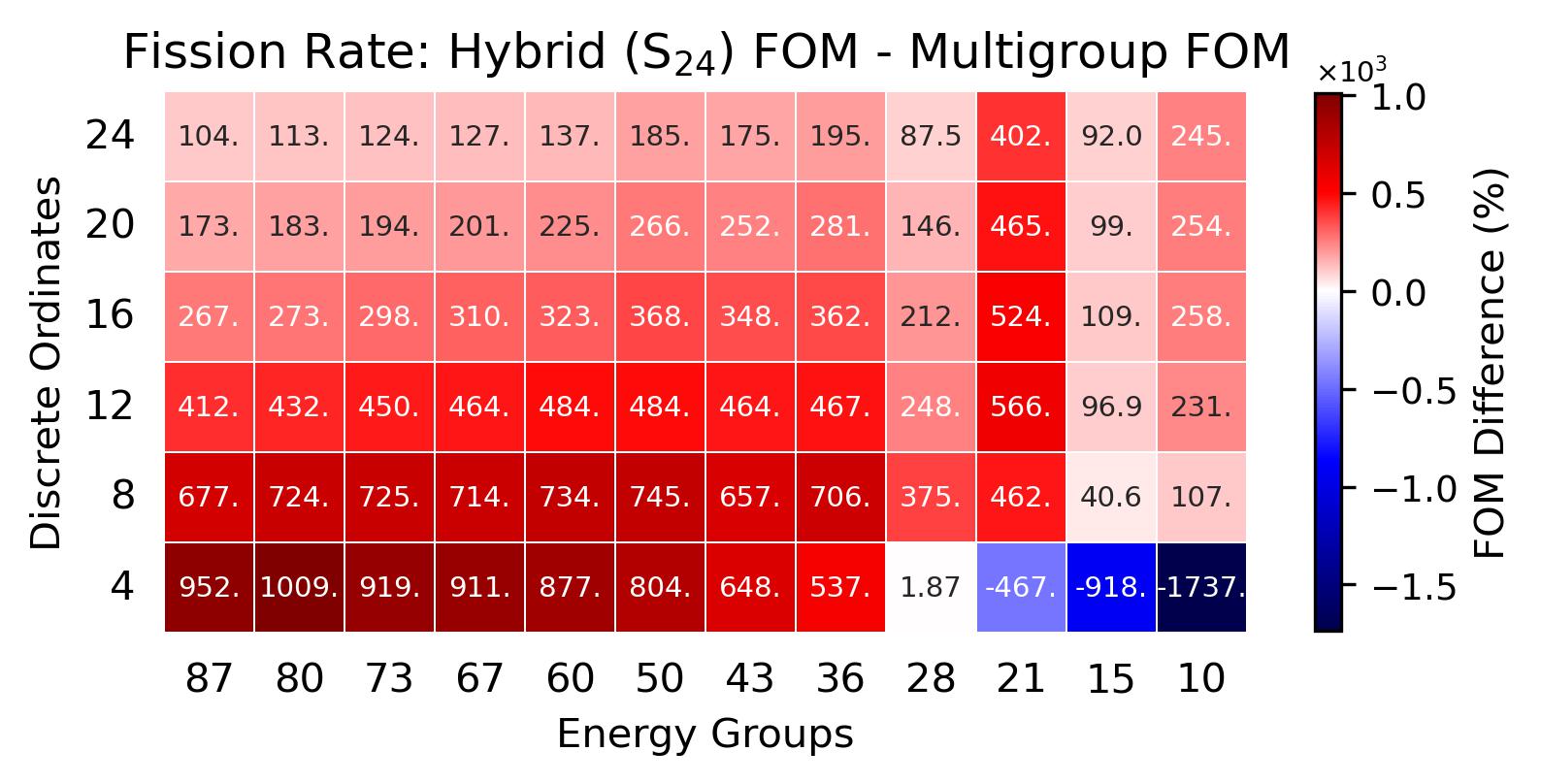} \label{fig:lattice_fission_fom_02}} 
    \subfloat[]{\centering \includegraphics[width=0.49\textwidth]{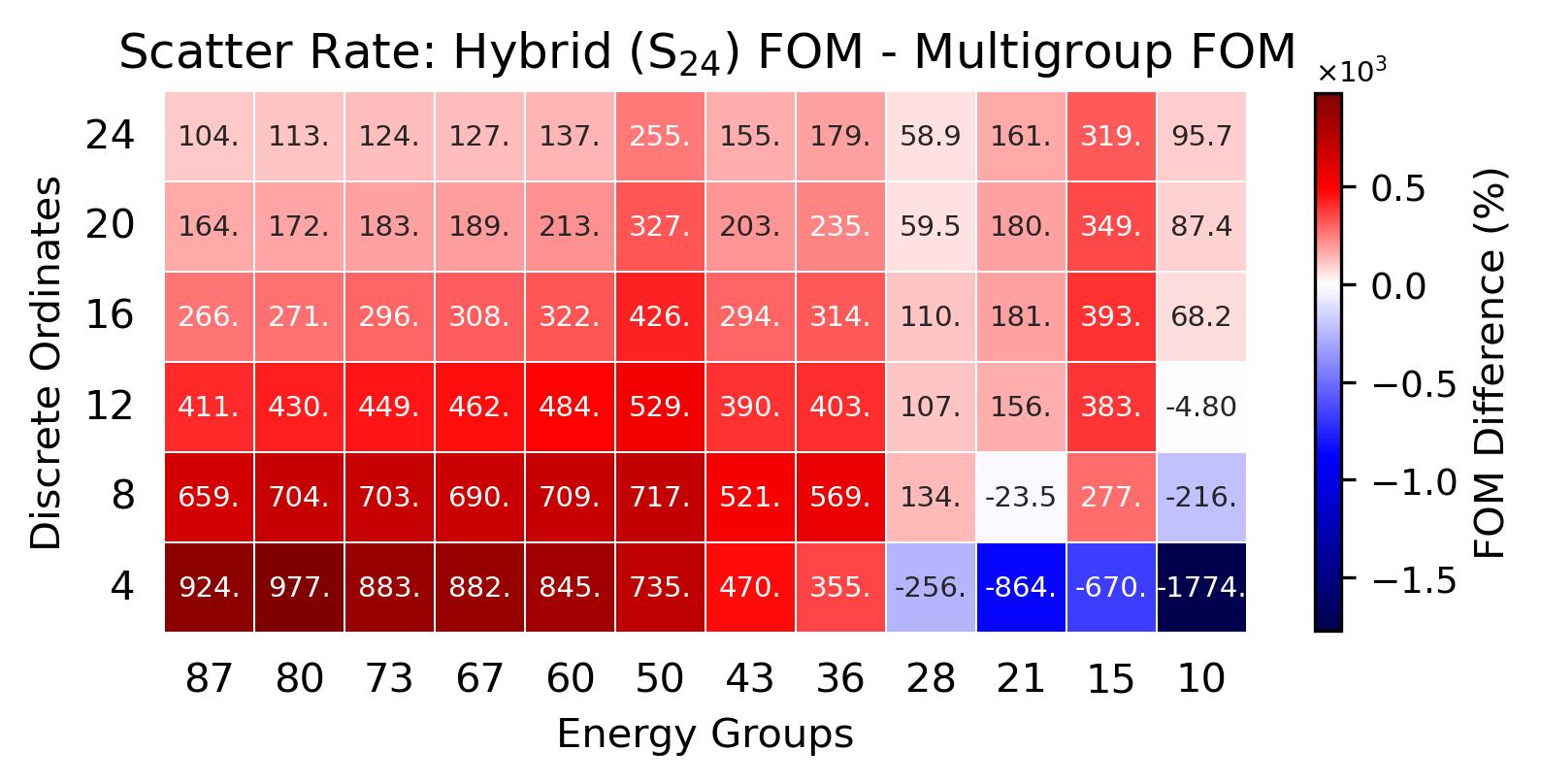} \label{fig:lattice_scatter_fom_02}} \\
    \subfloat[]{\centering \includegraphics[width=0.49\textwidth]{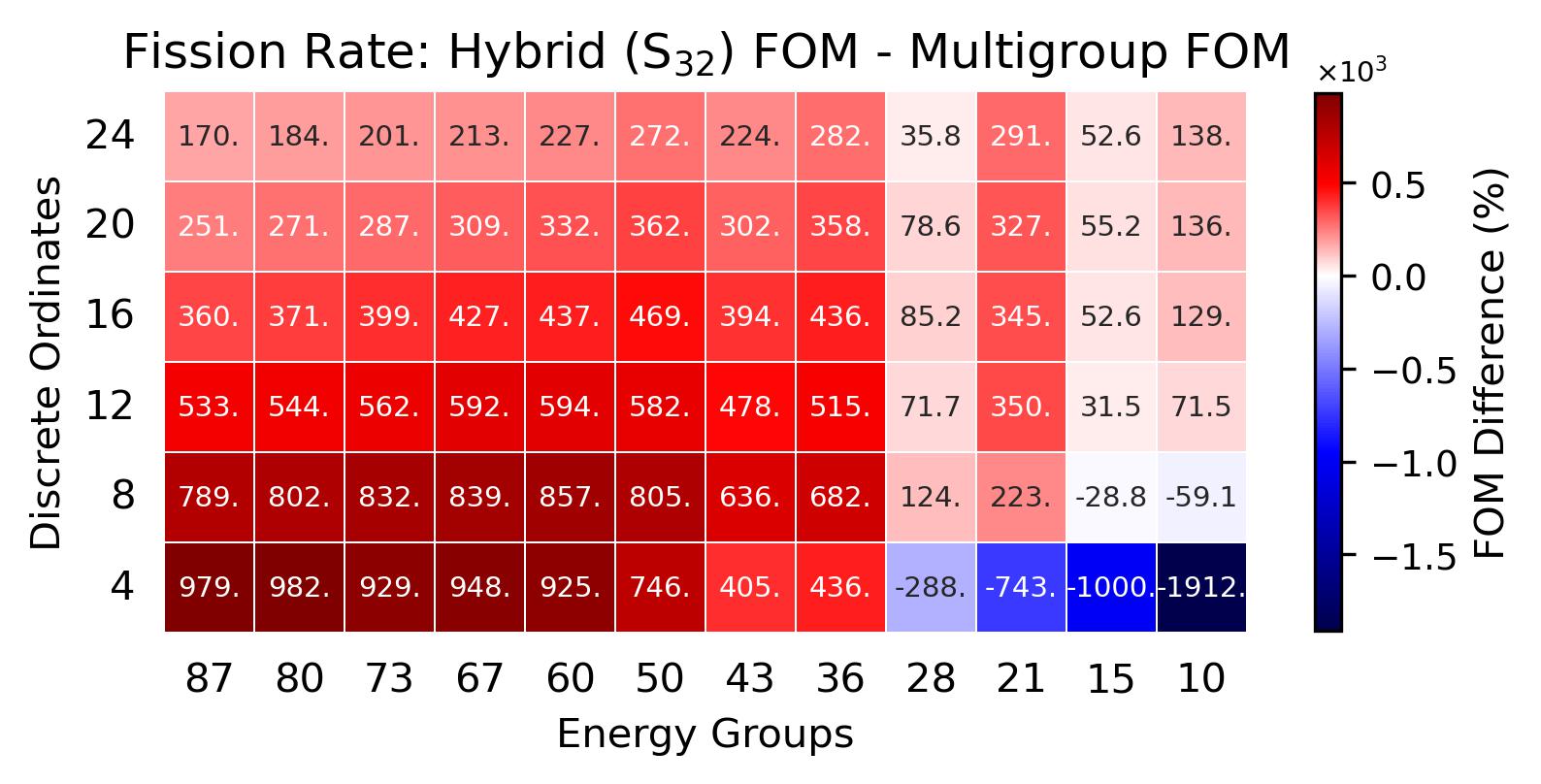} \label{fig:lattice_fission_fom_01}} 
    \subfloat[]{\centering \includegraphics[width=0.49\textwidth]{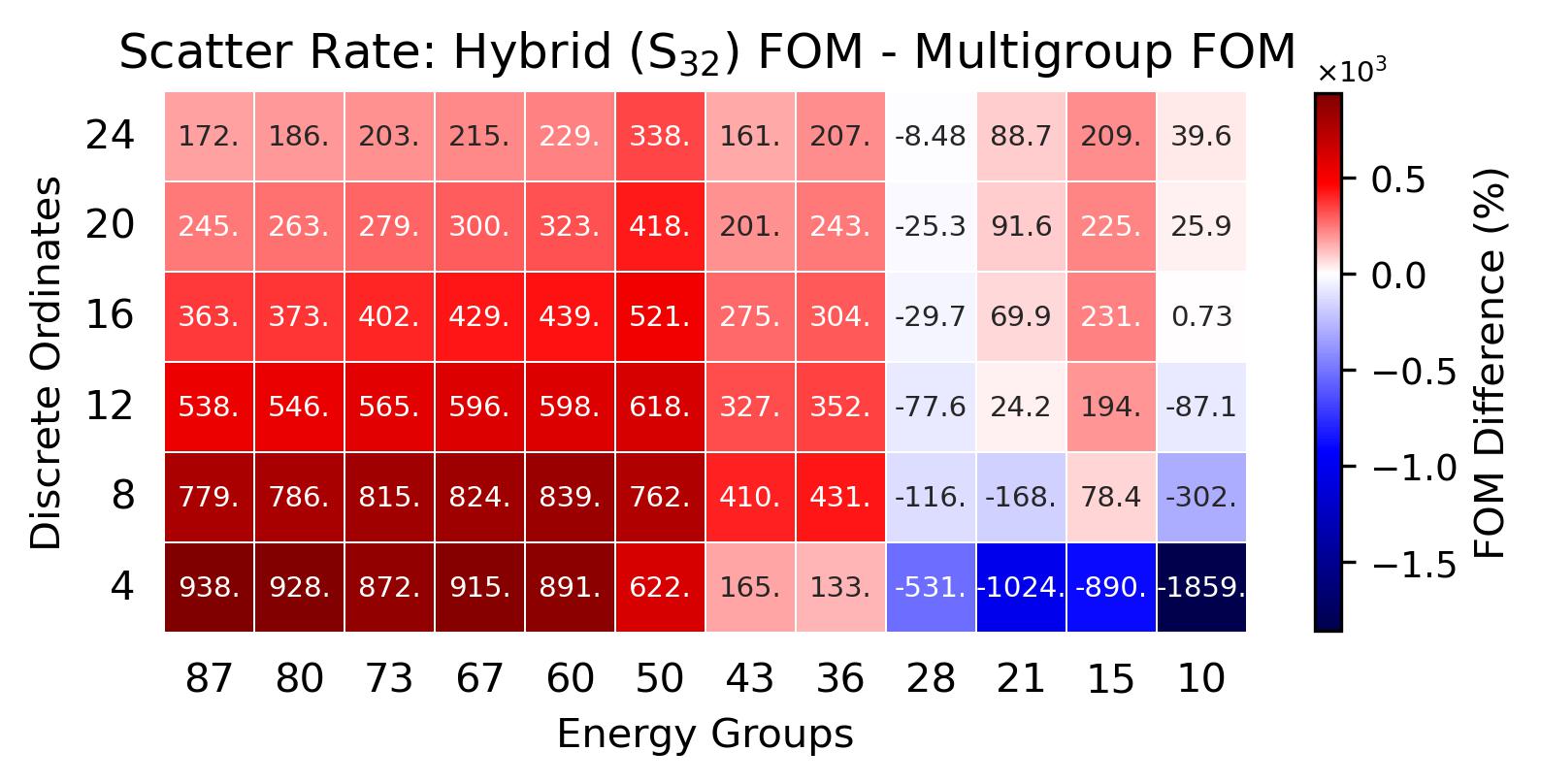} \label{fig:lattice_scatter_fom_01}}
    \caption{The figure of merit difference comparison for the AmBe lattice problem using an S$_{40}$ reference solution. Figure (a) uses the fission rate density RMSE with the S$_{24}$ uncollided hybrid model. Figure (b) shows the scattering rate density RMSE with the S$_{24}$ uncollided hybrid model. Figure (c) uses the fission rate density RMSE with the S$_{32}$ uncollided hybrid model. Figure (d) shows the scattering rate density RMSE with the S$_{32}$ uncollided hybrid model.}
    \label{fig:lattice_fom}
\end{figure}

The accuracy and elapsed time comparisons can combined into the figure of merit (FOM) metric, as described in Eq.~\eqref{eq:figure_of_merit}.
The FOM difference expressed as a percentage is shown in Fig.~\ref{fig:lattice_fom}. 
Figs. \ref{fig:lattice_fission_fom_02} and \ref{fig:lattice_scatter_fom_02} use the fission rate and scattering rate density errors for the S$_{24}$ uncollided hybrid method.
In most cases, the hybrid method outperforms the multigroup method, with the highest differences being in the high energy group and low angle models. 
The instances where the multigroup models outperform their hybrid counterparts are with the low angle and low energy group models. 
This observation is likely caused by the wall clock difference, as the hybrid model must iterate over the uncollided and corrector steps in addition to solving for the collided flux. 
This is also observed when using the S$_{32}$ uncollided hybrid models, as seen in Figs. \ref{fig:lattice_fission_fom_01} and \ref{fig:lattice_scatter_fom_01}.
As shown in Fig.~\ref{fig:lattice_time_01}, the hybrid method takes longer to converge than the S$_{24}$ hybrid model in Fig.~\ref{fig:lattice_time_02} in the low fidelity models. 
This correlates to the decrease in FOM performance seen in the low energy group models in Fig. \ref{fig:lattice_scatter_fom_01} as compared to Fig. \ref{fig:lattice_scatter_fom_02}.
For the higher fidelity models such as $G = 87$, $M = 12$, the difference between the hybrid and multigroup model in the S$_{32}$ comparison is over 100\% better than that for the S$_{24}$ model. 
This is accomplished while requiring more convergence time for the S$_{32}$ hybrid solution, which points to a cheap method to mitigate ray effects without significantly increasing convergence time.

\begin{figure}[!ht]
    \centering
    \subfloat[]{\centering \includegraphics[width=0.49\textwidth]{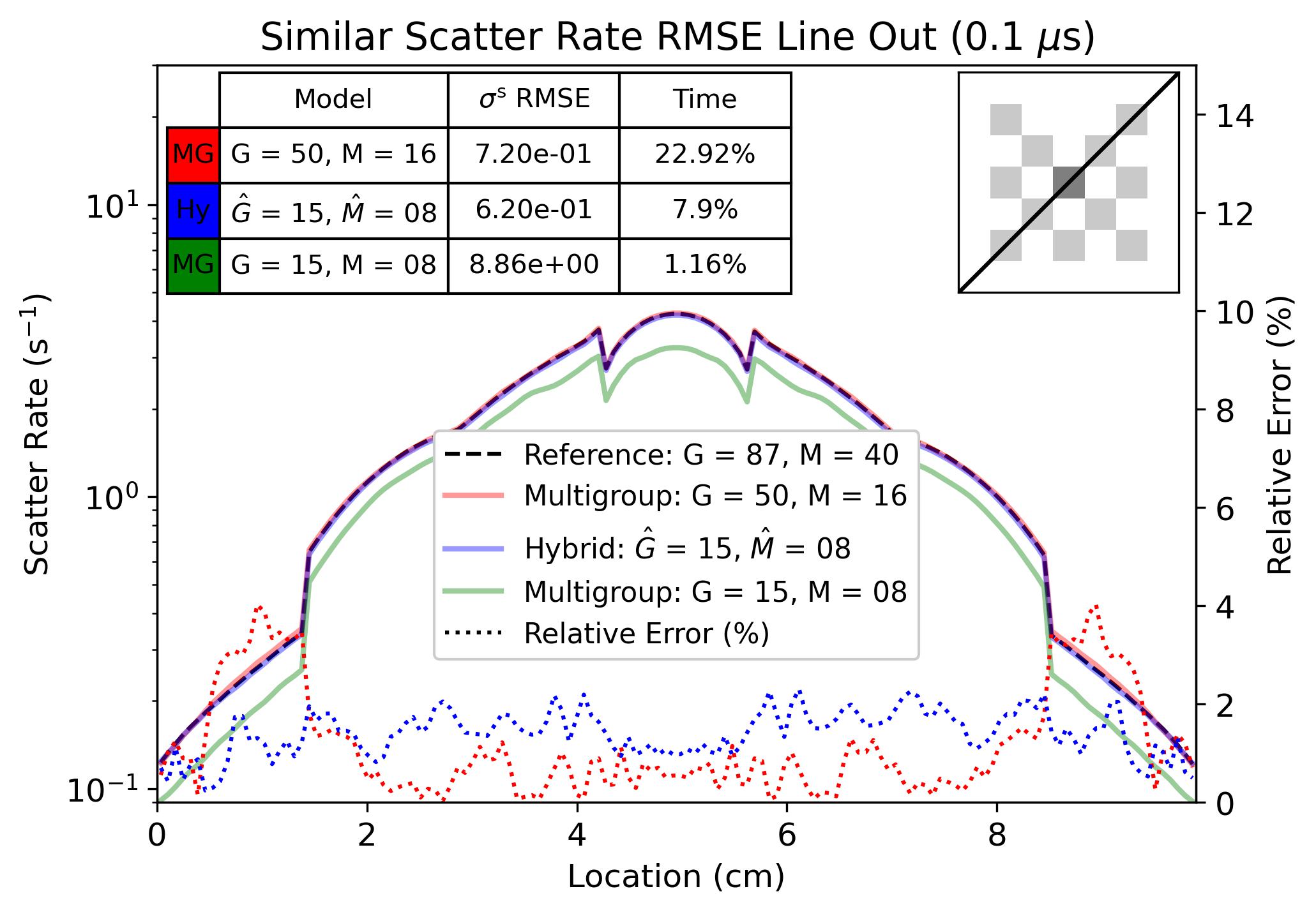} \label{fig:lattice_sim_err_line}}
    \subfloat[]{\centering \includegraphics[width=0.49\textwidth]{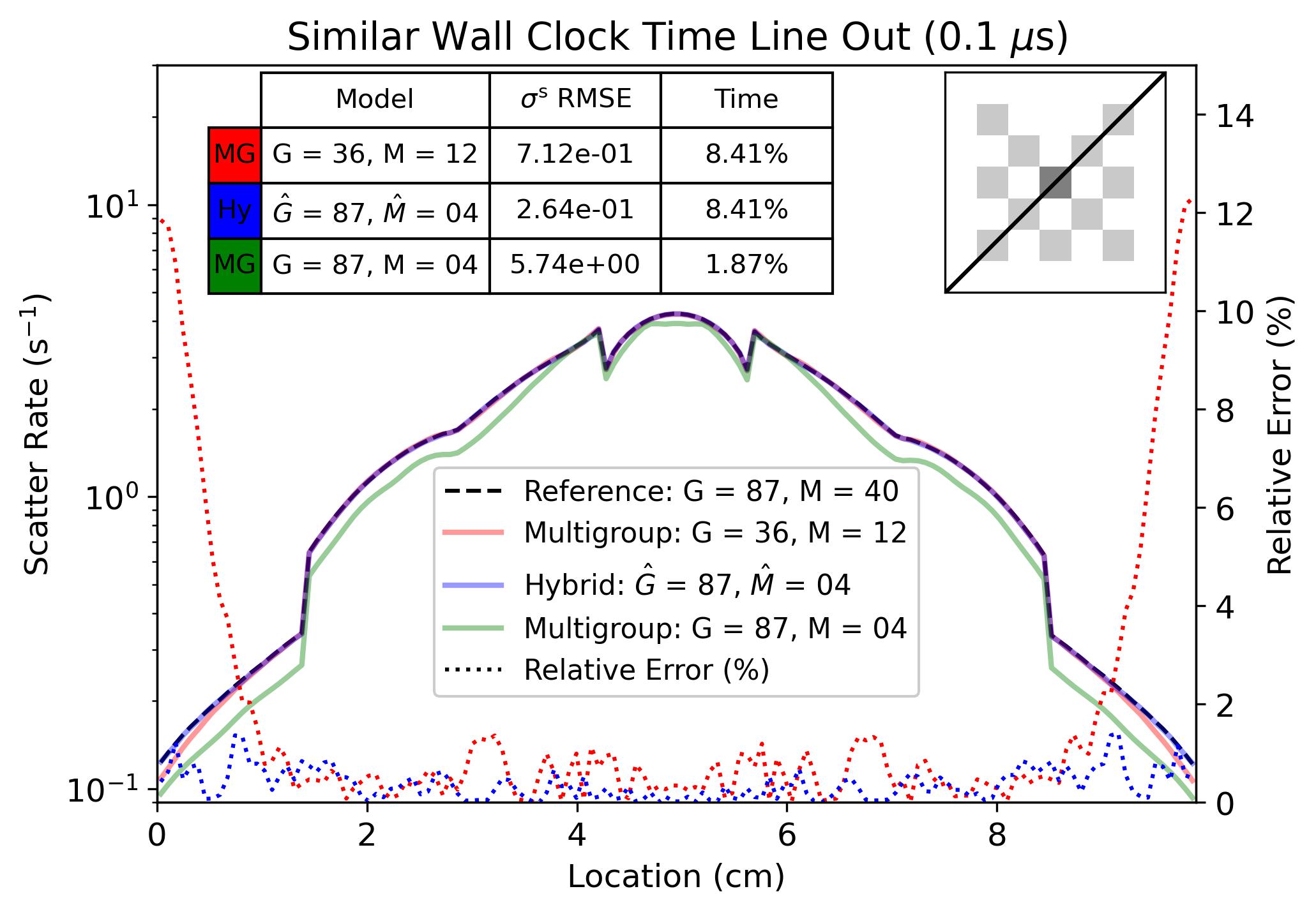} \label{fig:lattice_sim_time_line}}
    \caption{Line out comparisons of the hybrid and multigroup models for the AmBe lattice problem at the final time step. Figure (a) shows models with similar scattering rate density RMSE. Figure (b) shows models with similar wall clock convergence times.}
    \label{fig:lattice_line_outs}
\end{figure}

The benefits of using the hybrid method over monolithic coarsening can also be shown for models with different numbers of energy groups and discrete ordinates. 
Fig.~\ref{fig:lattice_sim_err_line} shows the line outs for a hybrid model with $G = 87$, $\hG = 15$, $M = 24$, $\hM = 8$ that has a similar scattering rate density error as a multigroup model with $G = 50$, $M = 16$. 
The scattering rate is similar between the hybrid and multigroup models but the hybrid model is able to converge in about a third of the time needed for the multigroup to converge. 
In addition, the multigroup model using the same numbers of energy groups and discrete ordinates as the collided hybrid model ($G = 15$, $M = 8$) is included. 
While this multigroup model converged about seven times faster than the hybrid model, the hybrid model was significantly more accurate in terms of scattering rate.

\begin{figure}[!ht]
    \centering
    \subfloat[]{\centering \includegraphics[width=0.495\textwidth]{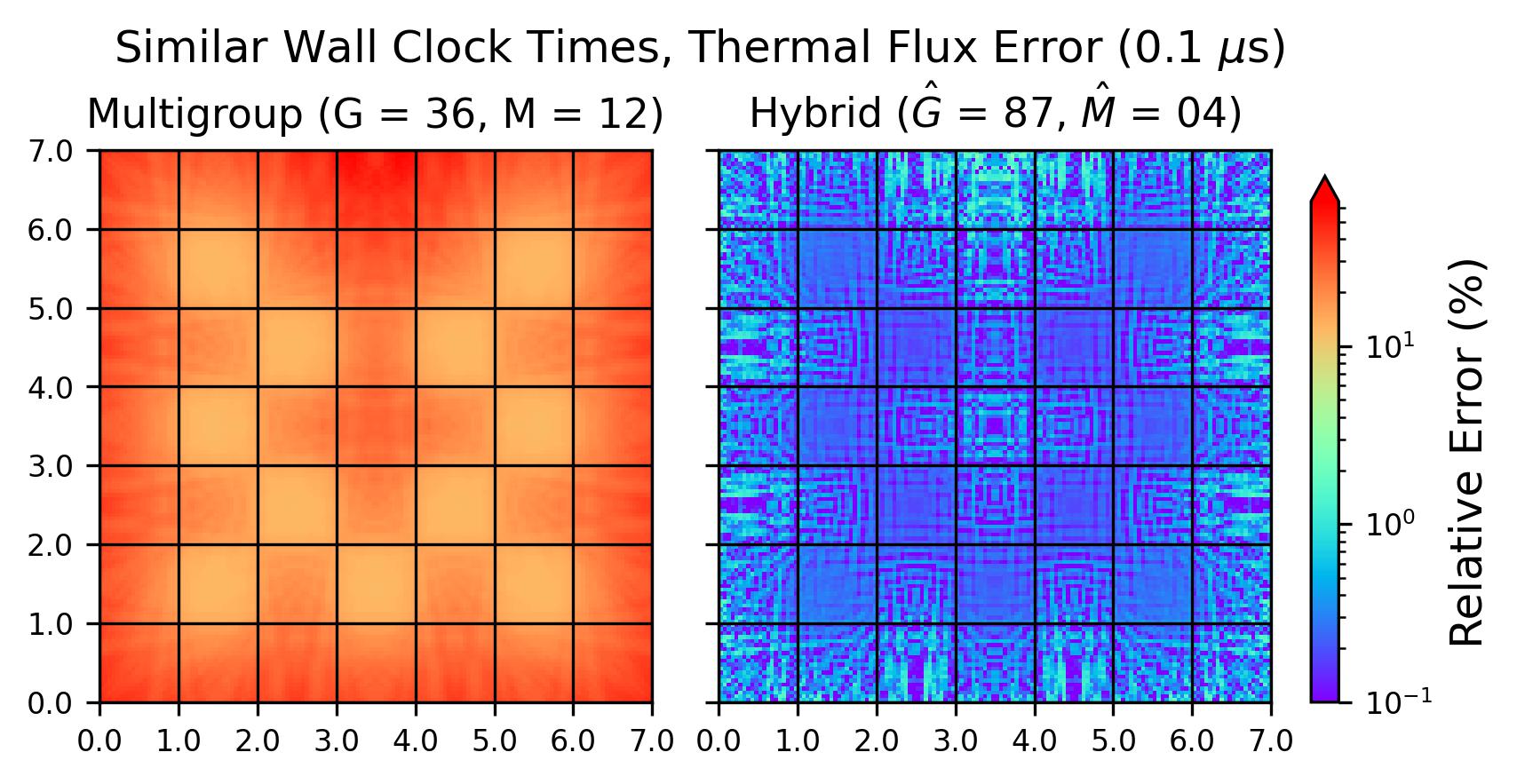} \label{fig:lattice_sim_time_thermal}} 
    \subfloat[]{\centering \includegraphics[width=0.495\textwidth]{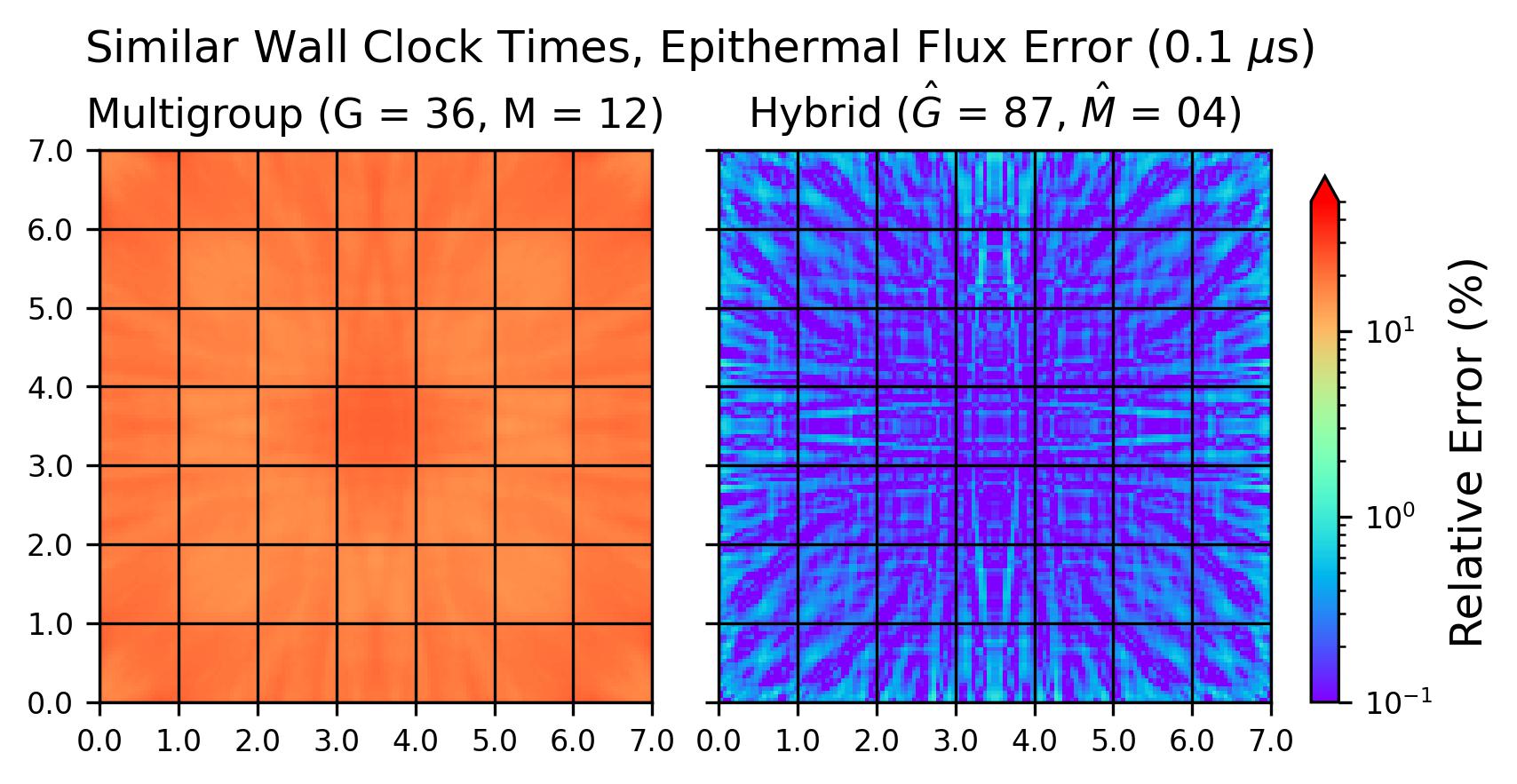} \label{fig:lattice_sim_time_epithermal}} \\
    \subfloat[]{\centering \includegraphics[width=0.495\textwidth]{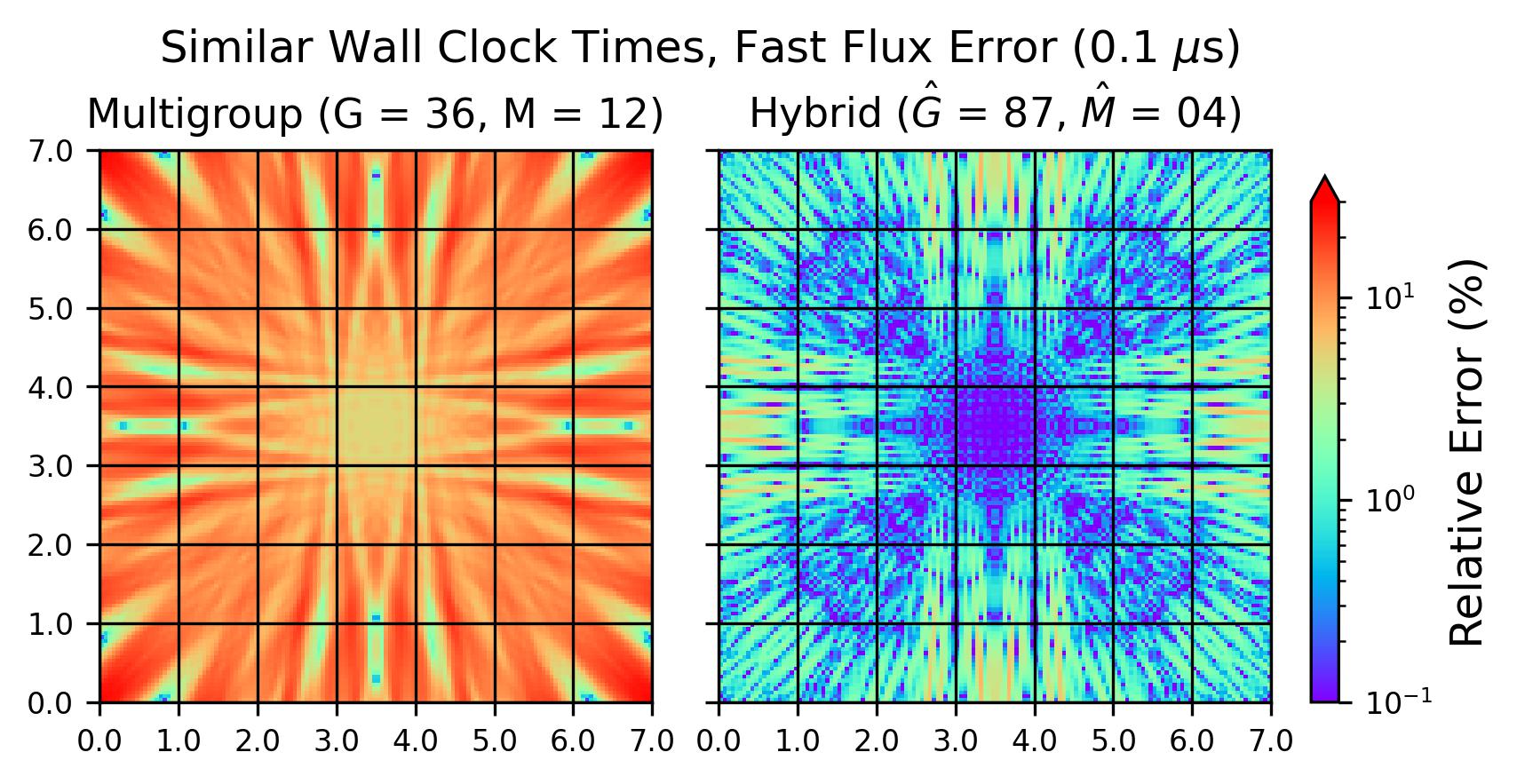} \label{fig:lattice_sim_time_fast}}
    \caption{The comparison of hybrid and multigroup models for the AmBe lattice problem that have similar wall clock times at the final time step. The relative error used an S$_{40}$ reference solution. Figure (a) is the error in the thermal region. Figure (b) is the error in the epithermal region. Figure (c) is the error in the fast region.}
    \label{fig:lattice_sim_time_energy}
\end{figure}

A similar line out comparison can be shown in terms of the wall clock convergence times, as seen in Fig.~\ref{fig:lattice_sim_time_line}. 
The hybrid model $G = 87$, $\hG = 87$, $M = 24$, $\hM = 4$ takes about the same amount of time to converge as the $G = 36$, $M = 12$ multigroup model.
While the scattering rate density error is similar in the middle of the problem, the multigroup model has a higher relative error towards the boundaries.
The multigroup model with the same numbers of energy groups and discrete ordinates is also compared, which shows an increase in convergence time but a decrease in accuracy.
This demonstrates that the hybrid method is more accurate than a multigroup model that requires similar convergence times. 
This is further exemplified by calculating the relative error between the reference flux and the hybrid and multigroup models, as shown in Fig.~\ref{fig:lattice_sim_time_energy} for each energy region.
For all three energy regions, thermal, epithermal, and fast, the hybrid model is significantly more accurate than the multigroup model while using a similar convergence time. 

\subsection{Double Chevron Problem}
\begin{figure}[!ht]
    \centering
    \includegraphics[width=0.4\textwidth]{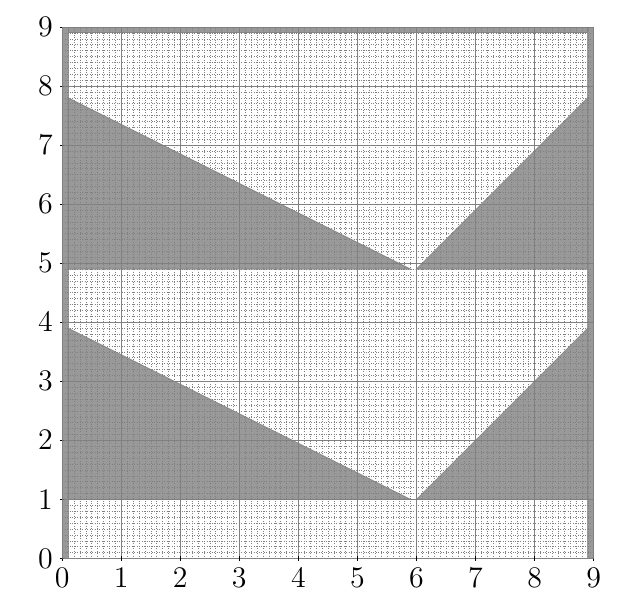}
    \caption{Setup for the double chevron problem. Natural uranium represents the chevrons (gray) with high density polyethylene in the center (white). A 14.1 MeV time-dependent source enters from the medium ($y = 0$ cm).}
    \label{fig:chevron_setup}
\end{figure}

The second problem uses a double chevron structure as shown in Fig. \ref{fig:chevron_setup}, where naturally enriched uranium was used for the chevrons and high density polyethylene (HDPE) was used for the scattering material.
A time-dependent source enters from the bottom of the medium ($y = 0$ cm) in the 14.1 MeV region.
The boundary source is represented as 
\begin{align} \label{eq:chevron_decay}
    b_{m, g=80} (x, y = 0, t) = 
    \begin{cases}
        0.1 \, t ,&\ t \in [0, 10] \\
        1.0 ,&\ t \in (10, 20] \\
        -0.5 \, t + 2 ,&\ t \in (20, 40] \\
        0 ,&\ t \in (40, 50]
    \end{cases}
\end{align}
where $t$ has units of $\mu$s with 50 evenly spaces time steps and a final time of $T$ = 50 $\mu$s.
The medium was $9 \times 9$ cm with vacuum boundaries and the number of spatial cells remained constant where $\Delta x = \Delta y = 0.1$ cm. 
The $G = 87$ energy group and S$_{32}$ angular grid was used for the reference solution.

\begin{figure}[!htb]
    \centering
    \subfloat[]{\centering \includegraphics[width=0.495\textwidth]{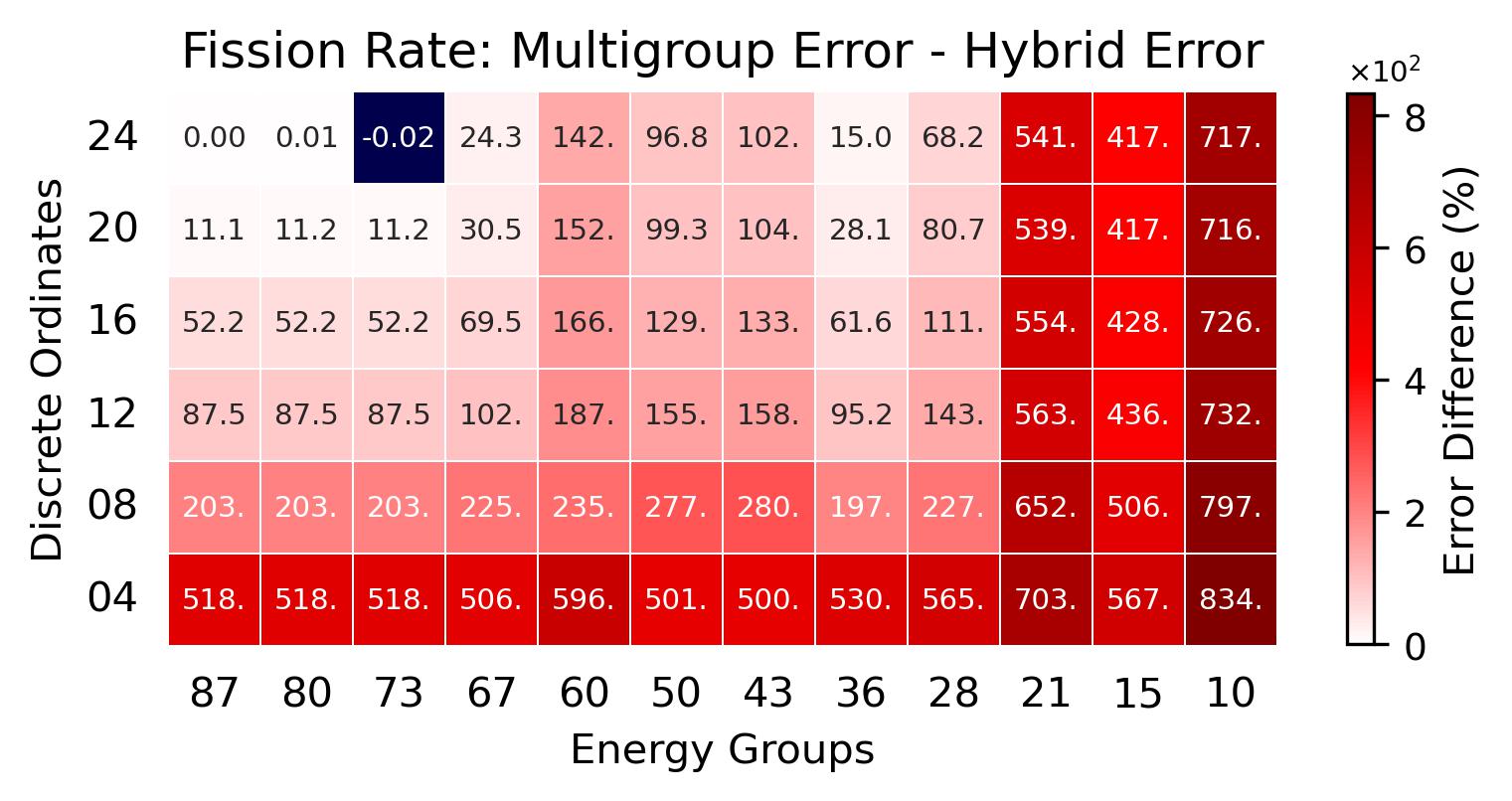} \label{fig:chevron_fission_error}}
    \subfloat[]{\centering \includegraphics[width=0.495\textwidth]{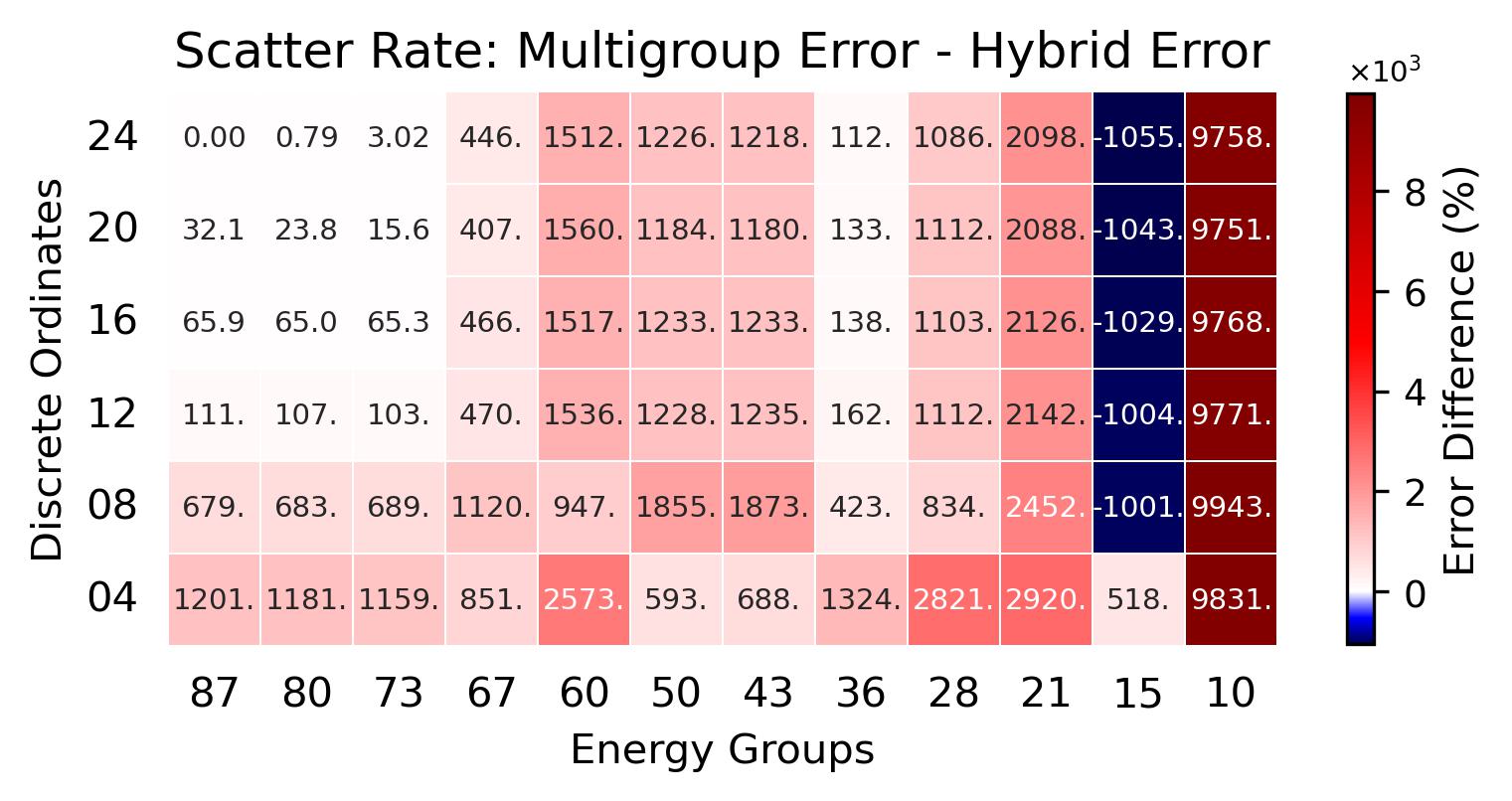} \label{fig:chevron_scatter_error}} \\
    \subfloat[]{\centering \includegraphics[width=0.495\textwidth]{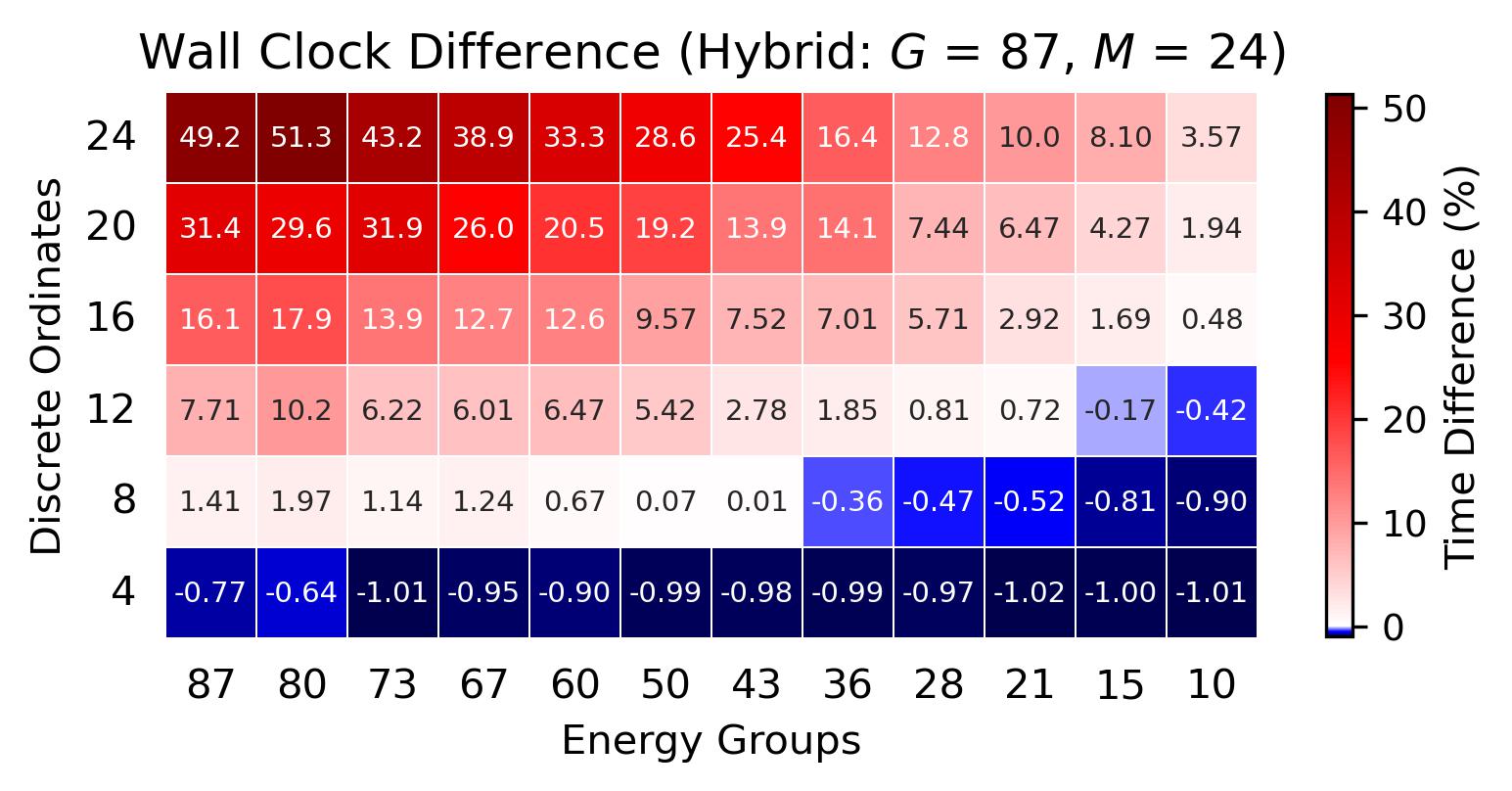} \label{fig:chevron_time}}
    \caption{Hybrid and multigroup model comparison for the double chevron problem. An S$_{32}$ reference solution is used with an S$_{24}$ uncollided hybrid solution in all instances. Figure (a) is the fission rate density RMSE percent difference. Figure (b) is the scattering rate density RMSE percent difference. Figure (c) compares the wall clock time difference.}
\end{figure}

The fission rate density RMSE difference between the multigroup and hybrid models is shown in Fig.~\ref{fig:chevron_fission_error}, where the uncollided hybrid model used an S$_{24}$ solution. 
In most instances, the hybrid model is more accurate than the multigroup model, with this difference becoming larger as the energy and angular grids are coarsened.
The only instance where the multigroup model outperforms the hybrid model is with the $G = 73$, $M = 24$, which is most likely due to the energy group coarsening scheme being suboptimal. 
This can also be seen when comparing the scattering rate density RMSE in Fig.~\ref{fig:chevron_scatter_error}.
The $G = 15$ models show that the multigroup is significantly more accurate than the hybrid models. 
Likewise, the $G = 36$ energy group structures for both the fission rate and scattering rate density errors also show the hybrid and multigroup models closer than higher fidelity models (i.e. $G = 43$). 
These trends have been observed with other problems using these group structures and should be considered side effects of the coarsening scheme.
Overall, the hybrid models tend to be more accurate than their equivalent multigroup models.

The wall clock time difference is shown in Fig.~\ref{fig:chevron_time} for the comparison of the hybrid and multigroup model convergence times. 
Each hybrid model used the S$_{24}$ uncollided solution and are shown as a percentage of the full model ($G = 87$, $M = 24$) wall clock time.
For the high fidelity models, the hybrid method converged up to 50\% faster than an equivalent multigroup model. 
The $G = 80$ and $M = 24$ result shows that the hybrid method is faster than the $G = 87$ and $M = 24$ model, which is likely due to a discrepancy in the number of iterations required for convergence.
This wall clock benefit disappears when coarsening the energy and angular grids, most likely caused by the additional $G \times M^2$  iterations for the uncollided and corrector steps at each time step.

\begin{figure}[!ht]
    \centering
    \subfloat[]{\centering \includegraphics[width=0.495\textwidth]{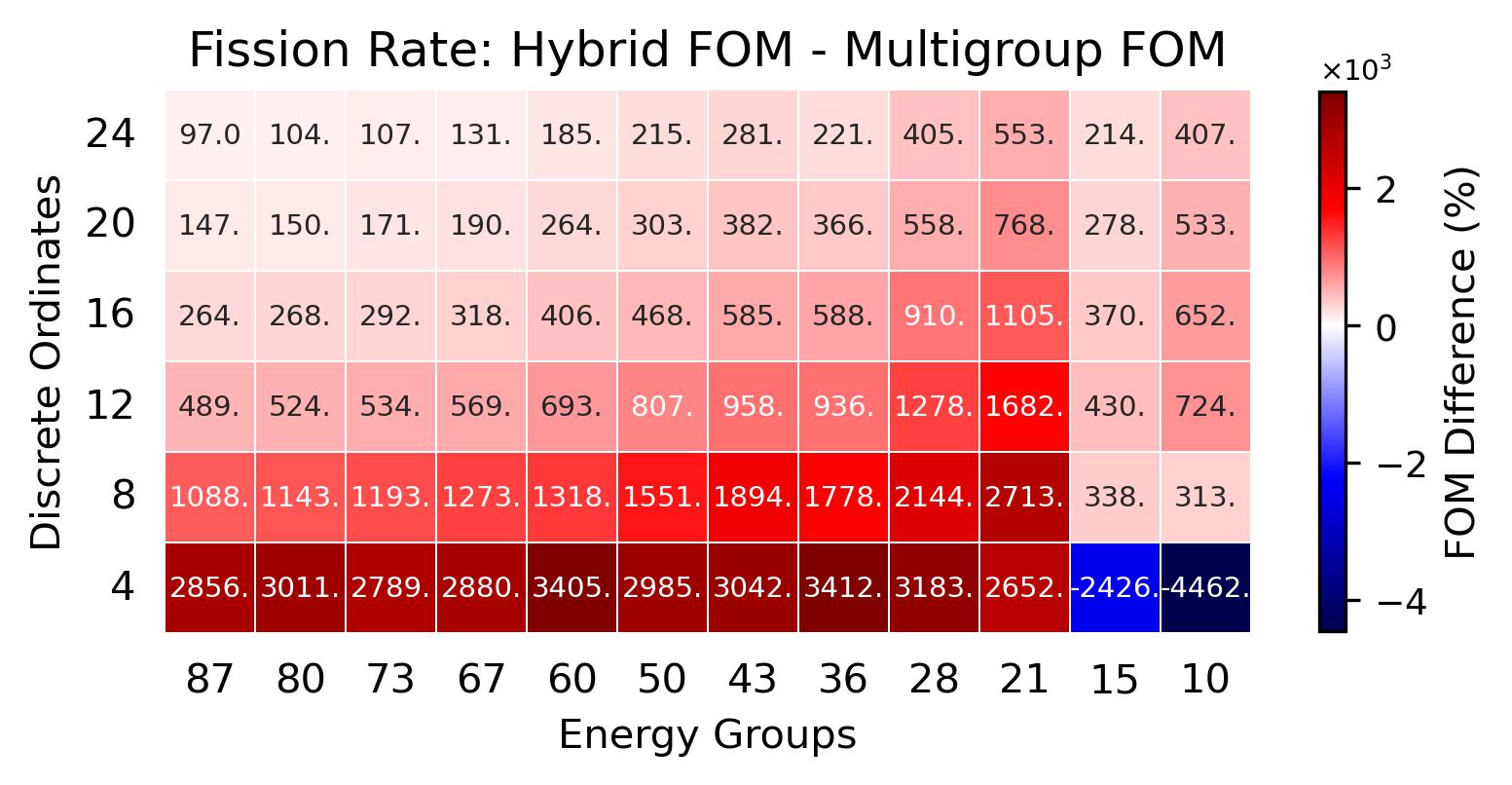} \label{fig:chevron_fission_fom}} 
    \subfloat[]{\centering \includegraphics[width=0.495\textwidth]{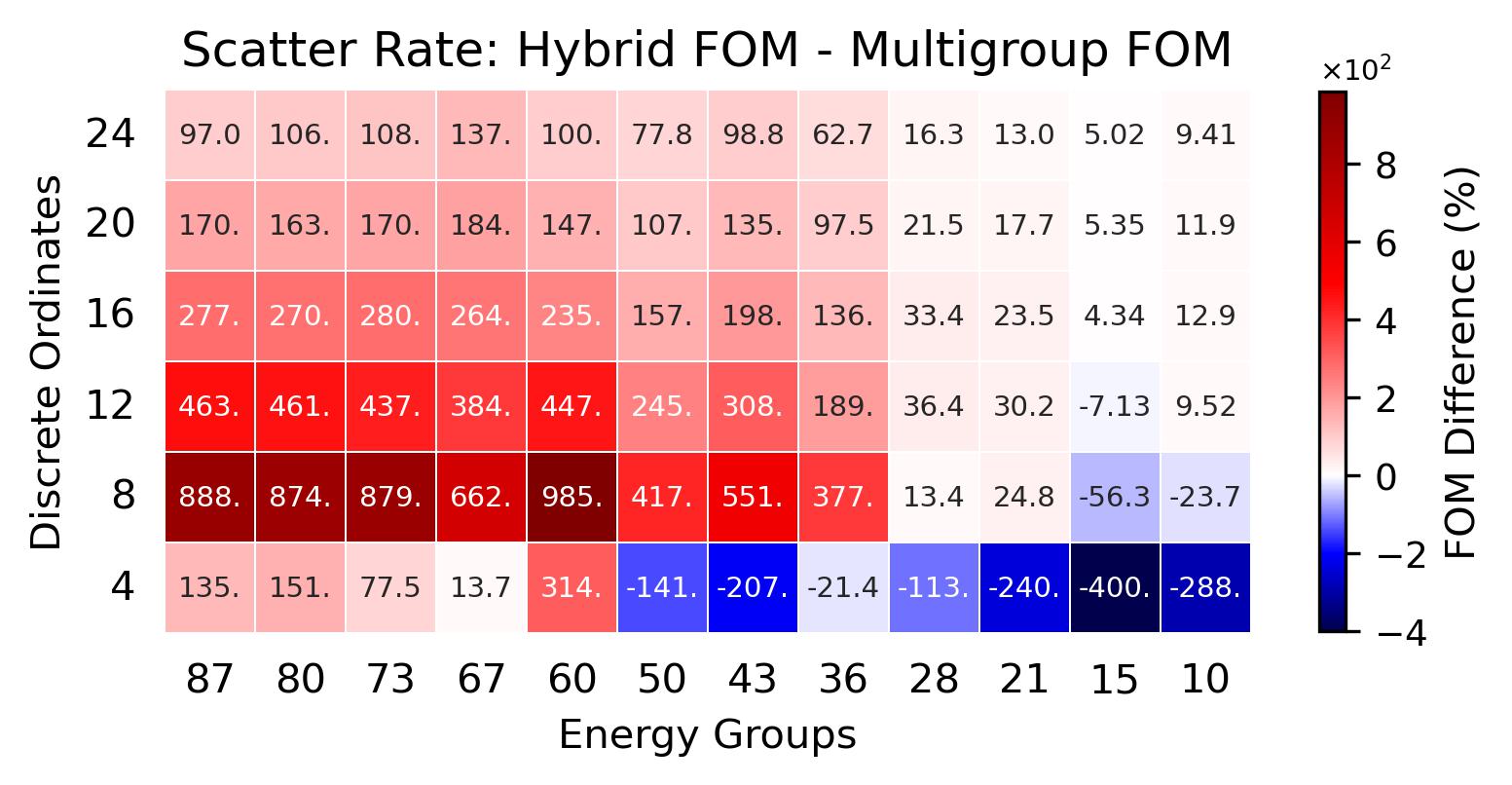} \label{fig:chevron_scatter_fom}}
    \caption{The figure of merit (FOM) difference comparison for the double chevron problem. The S$_{32}$ reference solution is used in addition to the S$_{24}$ uncollided solution. Figure (a) uses the fission rate density RMSE for the FOM calculation. Figure (b) uses the scattering rate density RMSE for the FOM calculation.}
    \label{fig:chevron_fom}
\end{figure}

The error metric can be combined with the wall clock time metric into the FOM based off of Eq.~\eqref{eq:figure_of_merit} and displayed in Fig.~\ref{fig:chevron_fom}.
The FOM percent difference that uses the fission rate density RMSE is shown in Fig.~\ref{fig:chevron_fission_fom}. 
In most cases, the hybrid model outperforms its equivalent multigroup model. 
The instances where the multigroup model is better are for the low fidelity cases, where the wall clock time of the multigroup model is significantly faster than the hybrid wall clock time. 
This can also be observed when using the scattering rate density RMSE in the FOM calculation, as seen in Fig.~\ref{fig:chevron_scatter_fom}.
The multigroup models that outperform their equivalent hybrid models are more prevalent in these cases because the scattering rate is not significantly better in the low fidelity models in addition to the previously mentioned wall clock issue. 
Although the low fidelity models have a higher FOM, their hybrid model counterpart tends to be more accurate.

\begin{figure}[!ht]
    \centering
    \subfloat[]{\centering \includegraphics[width=0.49\textwidth]{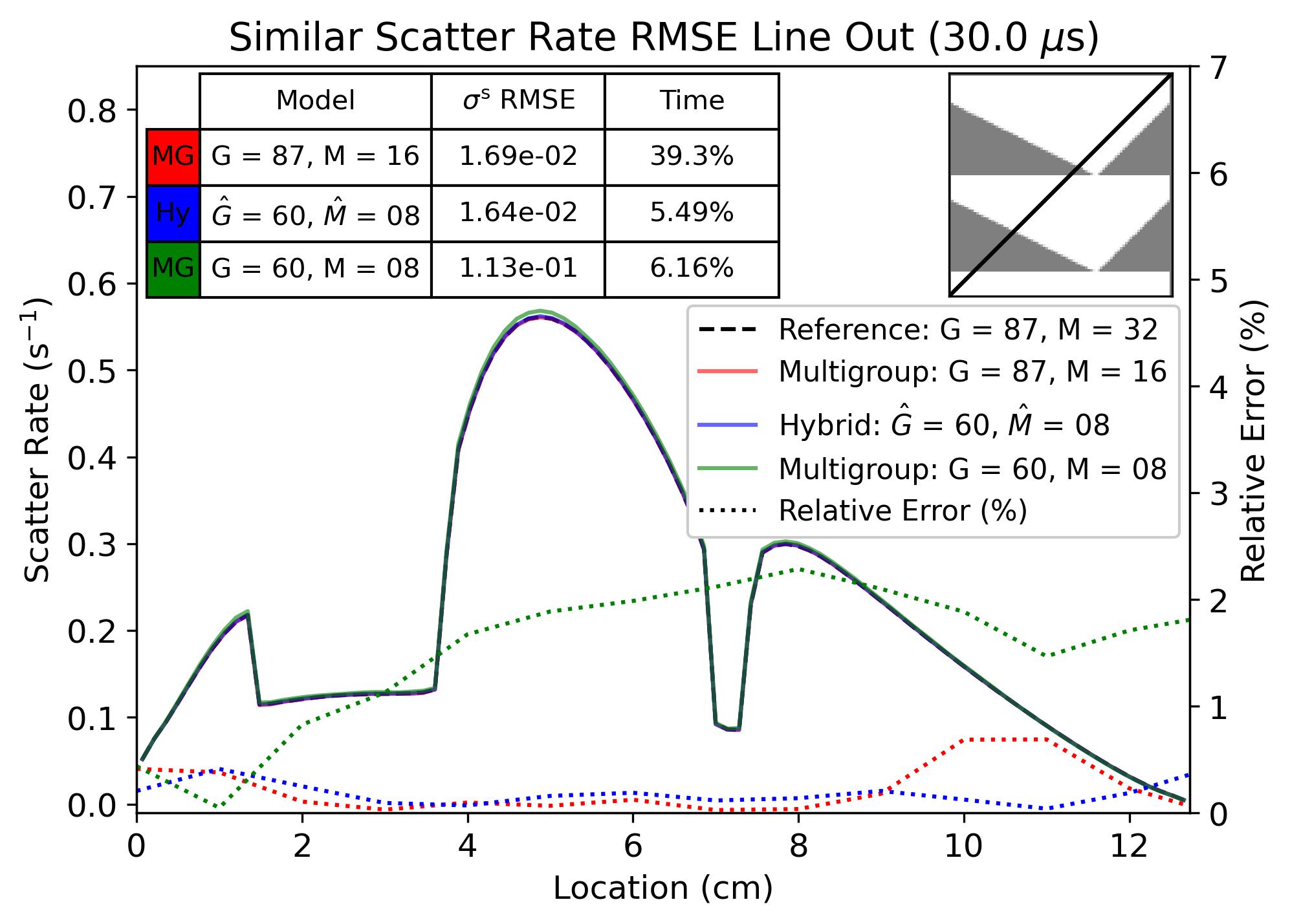} \label{fig:chevron_sim_err_line}} 
    \subfloat[]{\centering \includegraphics[width=0.49\textwidth]{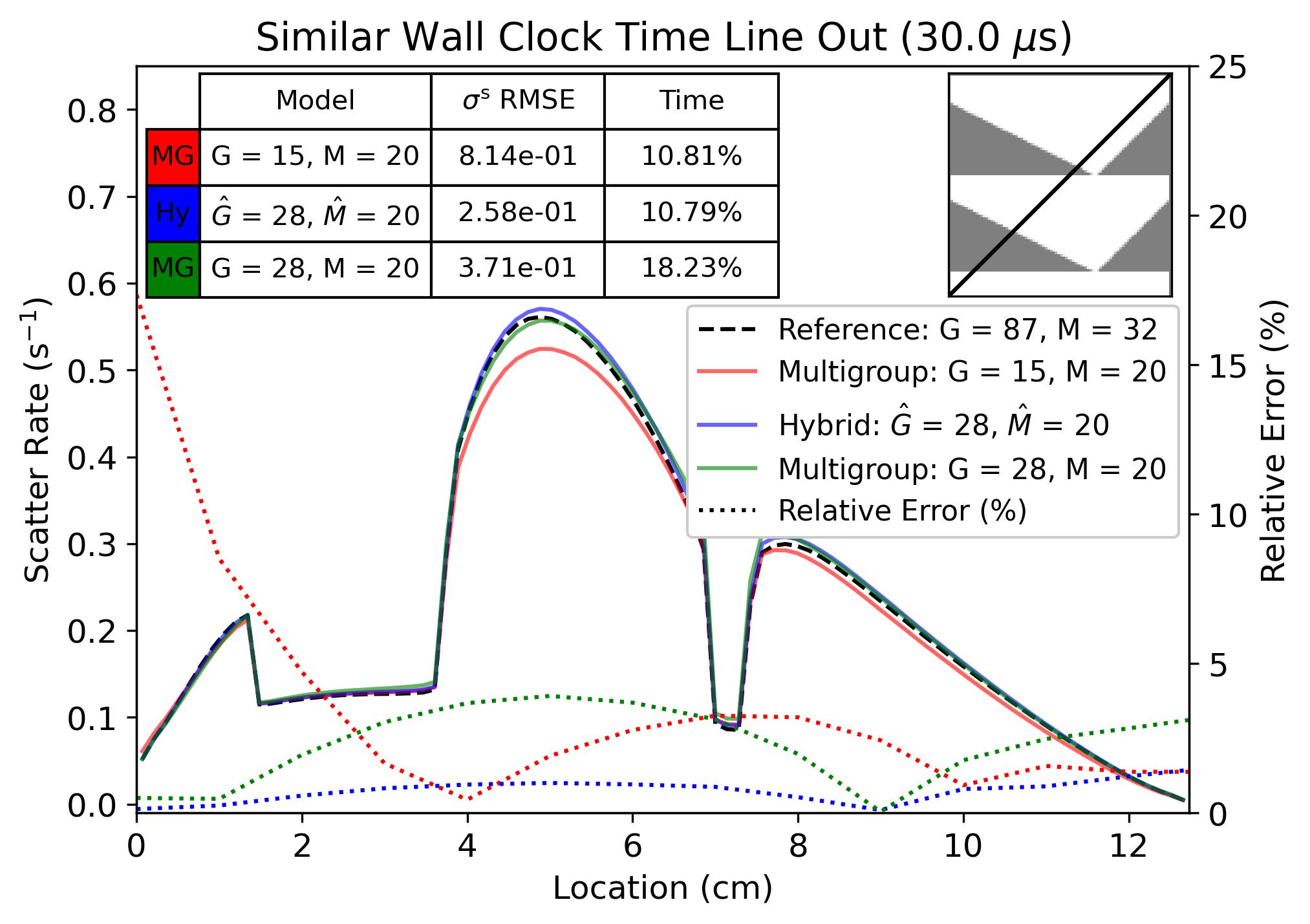} \label{fig:chevron_sim_time_line}}
    \caption{Line out comparisons of the hybrid and multigroup models for the double chevron problem at $t = 30 \, \mu$s. Figure (a) shows models with similar scattering rate density RMSE. Figure (b) shows models with similar wall clock convergence times.}
    \label{fig:chevron_line_outs}
\end{figure}

The benefits of using the hybrid method over monolithic coarsening methods can be shown for models with different numbers of energy groups and discrete ordinates. 
For instance, the hybrid model with $G = 87$, $\hG = 60$, $M = 24$, $\hM = 8$ has a similar scattering rate density RMSE as the $G = 87$, $M = 16$ multigroup model, as displayed by the scattering rate line out in Fig.~\ref{fig:chevron_sim_err_line}. 
This is shown at $t = 30 \, \mu$s, where the 14.1 MeV boundary source is decaying, although this trend can be observed at all time steps. 
The hybrid model was able to achieve an accuracy close to the multigroup model while converging over seven times faster. 
An additional multigroup model was included that used the same collided energy groups and discrete ordinates as the hybrid model ($G = 60$, $M = 8$) but required more wall clock time in addition to being less accurate. 

\begin{figure}[!ht]
    \centering
    \subfloat[]{\centering \includegraphics[width=0.495\textwidth]{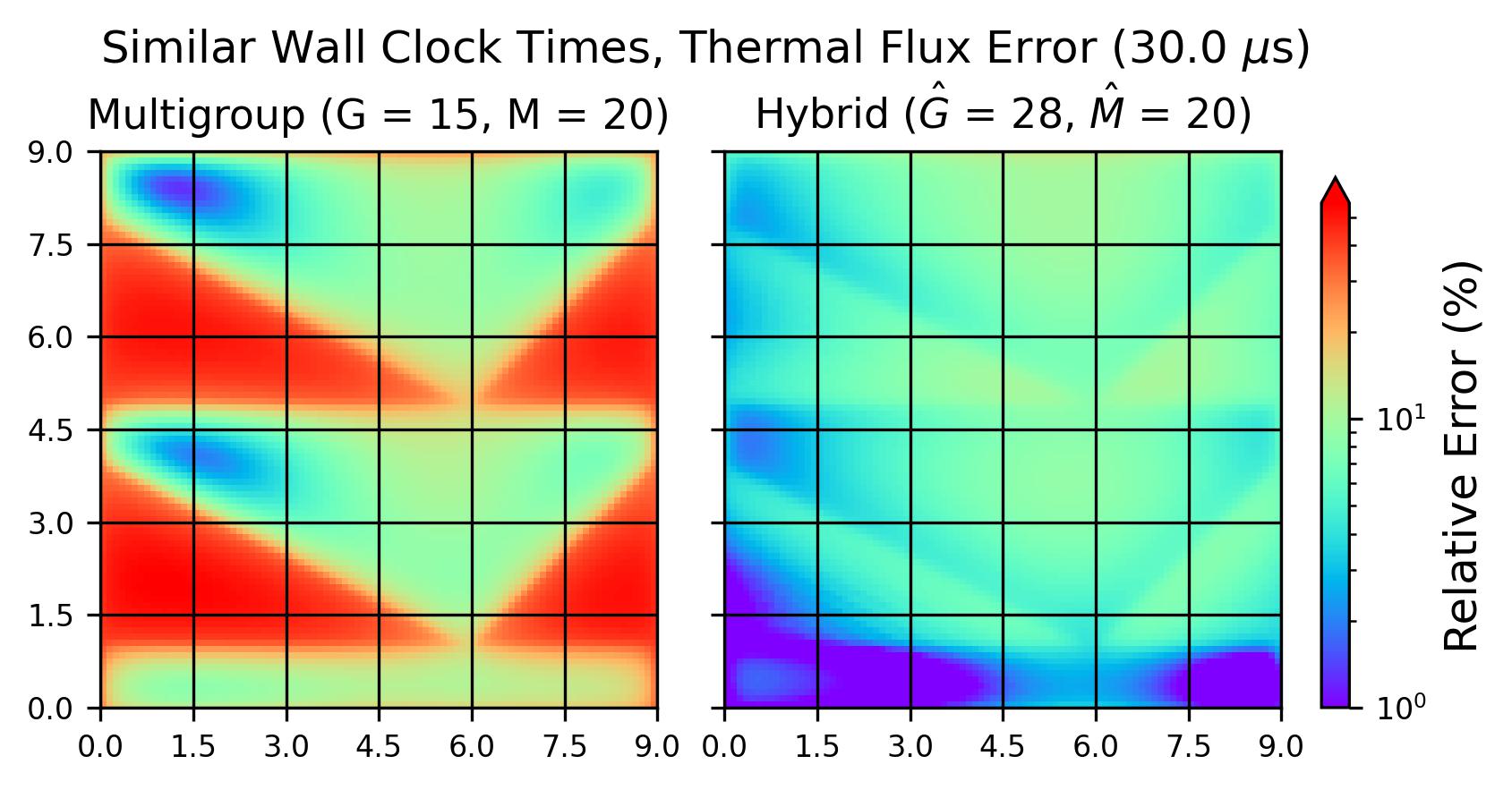} \label{fig:chevron_sim_time_thermal}} 
    \subfloat[]{\centering \includegraphics[width=0.495\textwidth]{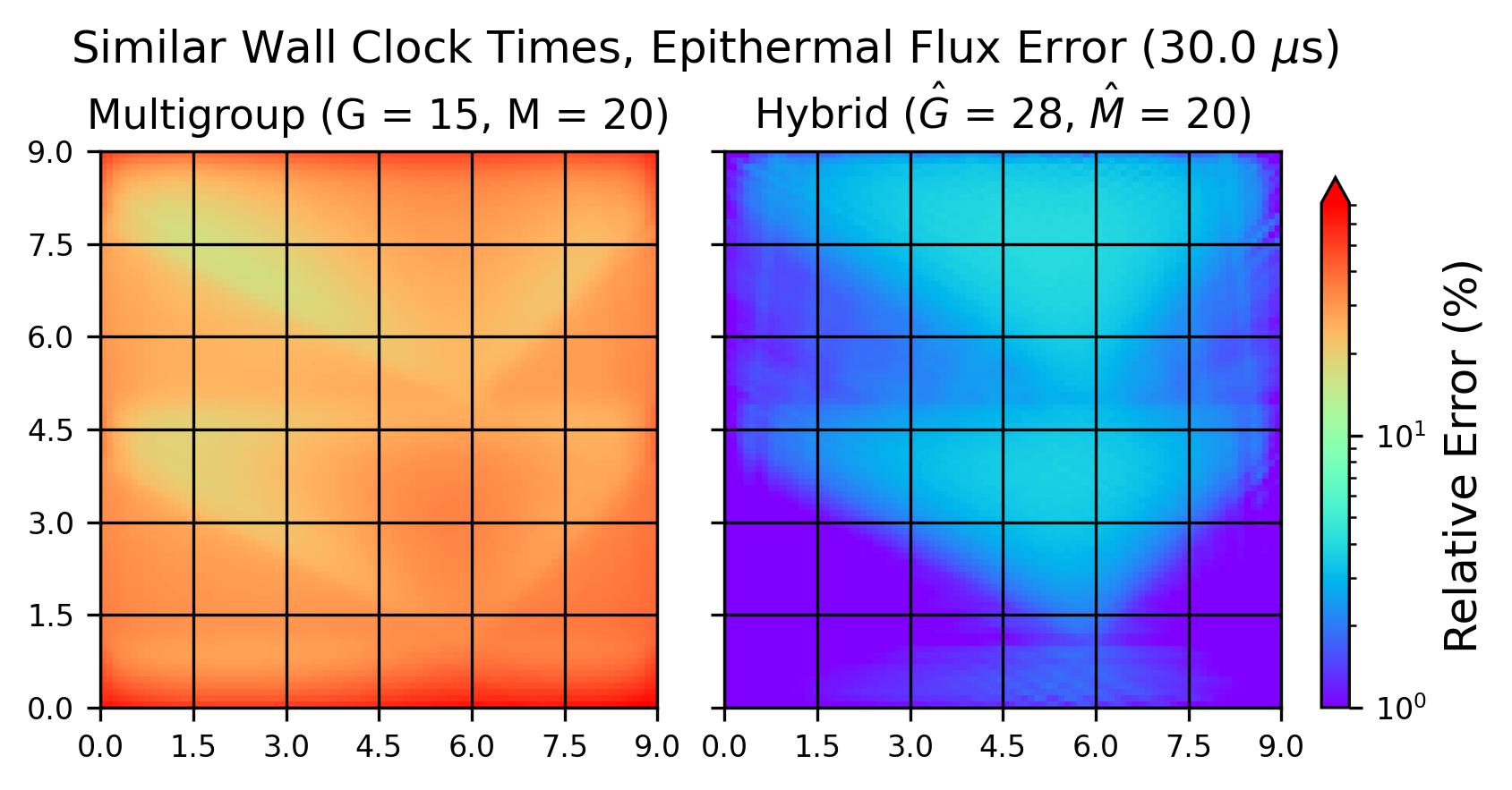} \label{fig:chevron_sim_time_epithermal}} \\
    \subfloat[]{\centering \includegraphics[width=0.495\textwidth]{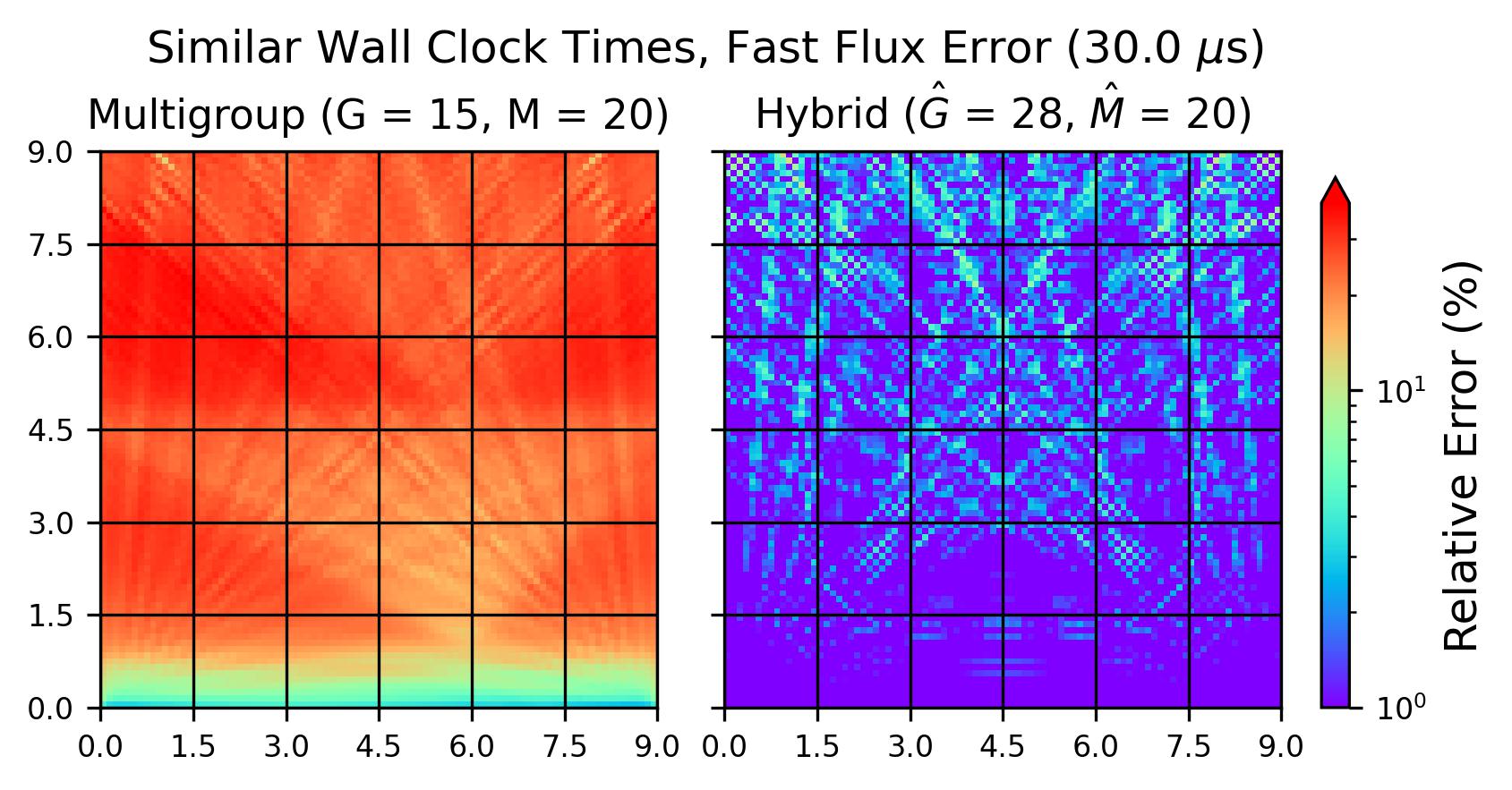} \label{fig:chevron_sim_time_fast}}
    \caption{The comparison of hybrid and multigroup models for the double chevron problem that have similar wall clock times at $t = 30 \, \mu$s. The relative error used an S$_{32}$ reference solution. Figure (a) is the error in the thermal region. Figure (b) is the error in the epithermal region. Figure (c) is the error in the fast region.}
    \label{fig:chevron_sim_time_energy}
\end{figure}

Hybrid and multigroup models with similar wall clock times can also be compared, as seen by the scattering rate line out in Fig.~\ref{fig:chevron_sim_time_line}.
For this model, both the hybrid model ($G = 87$, $\hG = 28$, $M = 24$, $\hM = 20$) and the multigroup model ($G = 15$ and $M = 20$) use about 10.8\% of the full model wall clock time. 
These models are compared to a scatter rate line out with the S$_{32}$ reference solution at $t = 30 \, \mu$s.
While they require the similar convergence times, the hybrid model is more accurate than multigroup model, especially in the middle region between the chevrons. 
A similar $G = 28$, $M = 20$ multigroup model shows similar accuracy to the hybrid model but requires addition time to converge. 
The accuracy of the hybrid and multigroup models is further shown by calculating the flux error for the thermal, epithermal, and fast energy regions at time step $t = 30 \, \mu$s, as shown in Fig.~\ref{fig:chevron_sim_time_energy}.
For each of the energy regions, the hybrid method is more accurate when compared to the S$_{32}$ reference solution.
The ray effects seen in the fast region of Fig.~\ref{fig:chevron_sim_time_fast} are mitigated by using the hybrid method over the multigroup method. 

\section{Conclusion and Future Work} \label{sec:conclusion}
The collision-based hybrid method splits the NTE into collided and uncollided components. 
The uncollided equation uses a high resolution grid to minimize the discretization error without adding a substantial burden on the required computational time.
This is combined with the low resolution collided equation solution to decrease the overall computational time.
The hybrid method is able to converge faster than monolithic grid coarsening schemes while returning lower error levels in most instances.
This paper expands upon previous work \cite{Whewell:2023} that introduced the collision-based hybrid method to one-dimensional multigroup neutron transport problems.
We have extended the hybrid method for two-dimensional multigroup neutron transport problems while using a higher order temporal discretization scheme and more optimal group structures.

Applying the hybrid method to two separate two-dimensional problems demonstrates the benefits of this method in terms of accuracy and wall clock time.
Except in isolated cases, the hybrid method is more accurate than an equivalent monolithic grid coarsening solution. 
The hybrid method also requires the use of less wall clock time than its multigroup counterpart, excluding the solutions with the coarsest angular and energy grids. 
When comparing solutions with similar wall clock times and different grid parameters, the hybrid results were more accurate. 
Similarly, models with different grid parameters and similar scattering rate density errors show that the hybrid results were able to achieve the same accuracy in less required time.
When combining the accuracy and elapsed time metrics into a figure of merit, the hybrid solutions were preferred to their equivalent multigroup solutions in most instances.

The collision-based hybrid method can be further explored in a number of different areas. 
The hybrid method can be expanded to three-dimensional neutron transport problems and combined with other acceleration techniques such as diffusion synthetic acceleration to further decrease convergence time.
The splitting of the temporal dimension into collided and uncollided equations and exploring correction techniques to remove coarse grid errors \cite{Crockatt:2020} can be investigated.
Ray effects can be an issue with two-dimensional discrete ordinates problems when not handled correctly \cite{Lewis:1993, Frank:2020}.
While there are different techniques for ray effect mitigation, the application to a coarse uncollided angular grid could improve wall clock times without affecting accuracy, decreasing the number of iterations for the uncollided and corrector steps.
Lastly, the hybrid method requires large amounts of memory. 
While this does not inhibit solving two-dimensional time-dependent neutron transport problems, it requires a higher memory footprint than monolithic grid coarsening. 
This is a research area than can be explored, by using an approach such as an on-the-fly cross section coarsening scheme to prevent storing both collided and uncollided cross sections.



\section*{Acknowledgments}
This work is supported by the Center for Exascale Monte-Carlo Neutron Transport (CEMeNT) a PSAAP-III project funded by the Department of Energy, grant number DE-NA003967.

\bibliographystyle{style/ans_js}
\bibliography{bibliography}

\newpage
\appendix
\section{Algorithms for the Multigroup and Hybrid Methods} \label{sec:algorithms}
\begin{algorithm}[H]
    \caption{TR-BDF2 time discretization for multigroup, discrete ordinates equation.}
    \label{alg:tr-bdf2}
    \setstretch{1.4}
    \begin{algorithmic}[1]
        \Require $\sig[g]{t}$, $\sig[g' \to g]{s}$, $\chi_g$, $\nu_{g'}$, $\sig[g' \to g]{f}$, $v_g$, $q_{m,g}$  \Comment{Material Properties}
        
        \Require $f$ \Comment{Initial Conditions}
        
        \Require $h$, $\bsOmega_m$, $w_m$, $\gamma$ \Comment{Discretization Parameters}
        
        \Require $\varepsilon_G, \varepsilon_M$ \Comment{Convergence Tolerances}
    
        \State $\psi_{g}\,^{(0)} \gets f_{g}$
        \State $\barpsi_{g}\,^{(0)} \gets \displaystyle\sum_{m=1}^M w_m f_{m,g}$
        
        \For{$n = 1, \dots, N$} \Comment{Iterate over Time Steps}
    
            \State $\sig[g]{*} \gets \displaystyle \frac{2}{\gamma \, v_{g} \, h} + \sig[g]{t}$
            
            \vspace{0.5em}
            
            \State $q_{m,g}^{*} \gets q_{m,g}^{(n-1+\gamma)} + \displaystyle \frac{2}{\gamma \, v_{g} \, h} \, \psi_{m,g}^{(n - 1)} + \sig[g' \to g]{s} \, \barpsi^{(n-1)} + \chi_{g} \nu_{g'} \sig[g' \to g]{f} \, \barpsi^{(n - 1)}_{g} $
            
             \hspace{35pt} $+ q_{m,g}^{(n - 1)} - \bsOmega \cdot \nabla \psi^{(n-1)}_{m,g} - \sig[g]{t} \, \psi^{(n - 1)}_{m,g}$ 
    
            \vspace{0.5em}
            \State $\psi_{m,g}^{(n-1+\gamma)} \gets$ Algorithm~\ref{alg:multigroup} with: \Comment{Crank-Nicolson Step} 
        
            \hspace{10pt} $\sig[g' \to g]{s} \gets \sig[g' \to g]{s}$, $\chi_{g}  \gets \chi_{g}$, $\nu_{g'} \gets \nu_{g'}$, $\sig[g' \to g]{f} \gets \sig[g' \to g]{f}$, $v_g \gets v_{g}$, $\sig[g]{*} \gets \sig[g]{*}$,
            
            \hspace{10pt} $q_{m,g}^{*} \gets q_{m,g}^{*}$, $\psi^{(n-1)} \gets \psi^{(n-1)}$, $\bsOmega_m \gets \bsOmega_{m}$, $w_m \gets w_{m}$, $\varepsilon_G \gets \varepsilon_G$, and $\varepsilon_M \gets \varepsilon_M$
    
            \vspace{0.5em}
            \State $\sig[g]{*} \gets \displaystyle \frac{2 - \gamma}{(1 - \gamma) \, v_{g} \, h} + \sig[g]{t}$
            
            \vspace{0.5em}
            \State $q_{m,g}^{*} \gets q_{m,g}^{(n)} + \displaystyle \frac{1}{\gamma \, (1 - \gamma) \, v_{g} \, h} \, \psi_{m,g}^{(n-1+\gamma)} - \frac{1 -\gamma}{\gamma \, v_{g} \, h} \, \psi_{m,g}^{(n-1)}$
    
            \vspace{0.5em}
            \State $\psi_{m,g}^{(n)} \gets$ Algorithm~\ref{alg:multigroup} with: \Comment{BDF2 Step} 
        
            \hspace{10pt} $\sig[g' \to g]{s} \gets \sig[g' \to g]{s}$, $\chi_{g} \gets \chi_{g}$, $\nu_{g'} \gets \nu_{g'}$, $\sig[g' \to g]{f} \gets \sig[g' \to g]{f}$, $v_g \gets v_{g}$, $\sig[g]{*} \gets \sig[g]{*}$,
            
            \hspace{10pt} $q_{m,g}^{*} \gets q_{m,g}^{*}$, $\psi^{(n-1)} \gets \psi^{(n-1+\gamma)}$, $\bsOmega_m \gets \bsOmega_{m}$, $w_m \gets w_{m}$, $\varepsilon_G \gets \varepsilon_G$, and $\varepsilon_M \gets \varepsilon_M$
        
        \EndFor
        
        \State \Return $\psi_{m,g} \,^{(N)}$
    \end{algorithmic} 
\end{algorithm}

\begin{algorithm}[H]
    \caption{Time step update of multigroup, discrete ordinates equation.}
    \label{alg:multigroup}
    \setstretch{1.4}
    \begin{algorithmic}[1]
        \Require $\sig[g' \to g]{s}$, $\chi_g$, $\nu_{g'}$, $\sig[g' \to g]{f}$, $v_g$  \Comment{Material Properties}
        
        \Require $\sig[g]{*}$, $q^{*}_{m,g}$, $\psi^{(n-1)}$ \Comment{Parameters Known from Previous Time Step}
        
        \Require $\bsOmega_m$, $w_m$ \Comment{Discretization Parameters}
        
        \Require $\varepsilon_G, \varepsilon_M$ \Comment{Convergence Tolerances}
        
        \State $\Delta_G \gets 1 + \varepsilon_G$, $j \gets 0$
        
        \State $\barpsi_{g}\,^{(0)} \gets \displaystyle\sum_{m=1}^M w_m \psi_{m,g}\,^{(n-1)}$
        
        \While{$\Delta_G$ $> \varepsilon_G$} \Comment{Outer Iteration ($j$)}
            \For{$g = 1, \dots, G$} \Comment{Loop over Groups}
                \State $\Tilde{Q}_{g} \gets \displaystyle\sum_{g'=1}^{g-1} \sig[g' \rightarrow g]{s} \barpsi_{g'} \,^{j+1} +
                \displaystyle\sum_{g' = g+1}^{G} \sig[g' \rightarrow g]{s} \barpsi_{g'} \,^{j}$ 
                
                \State $\Tilde{Q}_{g} \gets \Tilde{Q}_{g} + \chi_{g} \displaystyle\sum_{g'=1}^{g-1} \nu_{g'} \sig[g' \rightarrow g]{f} \barpsi_{g'} \,^{j+1} + \chi_{g} \displaystyle\sum_{g'= g+1}^{G} \nu_{g'} \sig[g' \rightarrow g]{f} \barpsi_{g'} \,^{j}$
                
                \vspace{0.5em}
                
                \State $\Delta_M \gets 1 + \varepsilon_M$, $\quad\ell \gets 0$
                
                \State $\barpsi_{g}\,^{j+1,0} \gets \barpsi_{g}\,^{j}$
    
                \While{$\Delta_M > \varepsilon_M$} \Comment{Source Iteration ($\ell$)}
                    \For{$m = 1, \dots, M$} \Comment{Loop over Angles}
    
                        \State $\Tilde{Q}_{m,g} \gets \Tilde{Q}_{g} + \sig[g \to g]{s}\barpsi_g\,^{j+1, \ell} + \chi_{g} \nu_{g} \sig[g \to g]{f}\barpsi_g\,^{j+1, \ell} + q^{*}_{m,g}$
    
                        \State $\psi_{m, g} \,^{j+1,\ell + 1} \gets \left(\bsOmega_m \cdot \nabla + \sig[g]{*}\right)^{-1} \Tilde{Q}_{m, g}$ \Comment{Transport Sweep}
                        
                    \EndFor
                    
                    \State $\barpsi_{g}\,^{j+1, \ell+1} \gets \displaystyle\sum_{m=1}^M w_m \psi_{m,g}\,^{j+1, \ell+1}$
                    
                    \State $\Delta_M$ $\gets \displaystyle \left| \left| \dfrac{\barpsi_{g} \,^{j+1, \ell + 1} - \barpsi_{g} \,^{j+1, \ell}}{\barpsi_{g} \,^{j+1, \ell + 1}} \right| \right|_{2}$
                    
                    \State $\ell \gets \ell + 1$
                \EndWhile
                
                \State $\psi_{m,g}\,^{j+1} \gets \psi_{m,g}\,^{j+1,\ell}\qquad\barpsi_{g}\,^{j+1} \gets \barpsi_{g}\,^{j+1, \ell}$
            
            \EndFor
            
            \State $\Delta_G$ $\gets \displaystyle \left| \left| \dfrac{\barpsi \,^{j + 1} - \barpsi \,^{j}}{\barpsi \,^{j + 1}} \right| \right|_{2}$ \Comment{$\barpsi \,^{j+1} = \left[ \, \barpsi_1 \,^{j+1}, \dots, \barpsi_G \,^{j+1} \, \right]$}
            
            \State $j \gets j + 1$
        \EndWhile 
        \State $\psi_{m,g}\,^{(n)} \gets \psi_{m,g}\,^{j},\qquad
        \barpsi_{g}\,^{(n)} \gets \barpsi_{g}\,^{j}$
        \State \Return $\psi_{m,g} \,^{(n)}$, $\barpsi_g \,^{(n)}$
    \end{algorithmic} 
\end{algorithm}

\begin{algorithm}[H]
    \caption{TR-BDF2 time discretization for collision-based hybrid multigroup, discrete ordinates equation.}
    \label{alg:hy-tr-bdf2}
    \begin{algorithmic}[1]
        \Require $\sig[g]{t}$, $\sig[g' \to g]{s}$, $\chi_{g}$, $\nu_{g'}$, $\sig[g' \to g]{f}$, $v_g$, $q_g$, \Comment{Uncollided Material Properties}
        
        \Require $\hsig[\hg]{t}$, $\hsig[\hg' \to \hg]{s}$, $\hat \chi_{\hg}$, $\hat \nu_{\hg'}$, $\hsig[\hg' \to \hg]{f}$, $v_{\hg}$ \Comment{Collided Material Properties}
        
        \Require $f$ \Comment{Initial Conditions}
        
        \Require $h$, $\bsOmega_m$, $w_{m}$, $\hat{\bsOmega}_{\hm}$, $\hw_{\hm}$, $\Delta E_g$, $\Delta \hE_{\hg}$, $\gamma$ \Comment{Discretization Parameters}
        
        \Require $\varepsilon_G, \varepsilon_M$ \Comment{Convergence Tolerances}
    
        \State $\psi_{g}\,^{(0)} \gets f_{g}$
        \State $\barpsi_{g}\,^{(0)} \gets \displaystyle\sum_{m=1}^M w_m f_{m,g}$
        
        \For{$n = 1, \dots, N$} \Comment{Iterate over Time Steps}
    
            \State $\sig[g]{*} \gets \displaystyle \frac{2}{\gamma \, v_{g} \, h} + \sig[g]{t}$ \qquad $\hsig[\hg]{*} \gets \displaystyle \frac{2}{\gamma \, v_{\hg} \, h} + \hsig[\hg]{t}$
            
            \vspace{0.5em}
            \State $q_{m,g}^{*} \gets q_{m,g}^{(n-1+\gamma)} + \displaystyle \frac{2}{\gamma \, v_{g} \, h} \, \psi_{m,g}^{(n - 1)} + \sig[g' \to g]{s} \, \barpsi^{(n-1)} + \chi_{g} \nu_{g'} \sig[g' \to g]{f} \, \barpsi^{(n - 1)}_{g} $
            
             \hspace{35pt} $+ q_{m,g}^{(n - 1)} - \bsOmega \cdot \nabla \psi^{(n-1)}_{m,g} - \sig[g]{t} \, \psi^{(n - 1)}_{m,g}$ 

            \vspace{0.5em}
            \State $\psi_{m,g}^{(n-1+\gamma)} \gets$ Algorithm~\ref{alg:hy-multigroup} with: \Comment{Crank-Nicolson Step} 
        
            \hspace{10pt} $\sig[g]{*} \gets \sig[g]{*}$, $\sig[g' \to g]{s} \gets \sig[g' \to g]{s}$, $\chi_{g} \gets \chi_{g}$, $\nu_{g'} \gets \nu_{g'}$, $\sig[g' \to g]{f} \gets \sig[g' \to g]{f}$, $v_g \gets v_{g}$, $q_{m,g}^{*} \gets q_{m,g}^{*}$,

            \hspace{10pt} $\hsig[\hg]{*} \gets \sig[\hg]{*}$, $\hsig[\hg' \to \hg]{s} \gets \hsig[\hg' \to \hg]{s}$, $\hat \chi_{\hg} \gets \hat \chi_{\hg}$, $\hat \nu_{\hg'} \gets \hat \nu_{\hg'}$, $\hsig[\hg' \to \hg]{f} \gets \hsig[\hg' \to \hg]{f}$, $v_{\hg} \gets v_{\hg}$,
            
            \hspace{10pt} $\psi^{(n-1)} \gets \psi^{(n-1)}$, $\bsOmega_m \gets \bsOmega_{m}$, $w_m \gets w_{m}$, $\hat{\bsOmega}_{\hm} \gets \hat{\bsOmega}_{\hm}$, $\hw_{\hm} \gets \hw_{\hm}$, 
            
            \hspace{10pt} $\Delta E_g \gets \Delta E_g$, $\Delta \hE_{\hg} \gets \Delta \hE_{\hg}$, $\varepsilon_G \gets \varepsilon_G$, and $\varepsilon_M \gets \varepsilon_M$
    
            \vspace{0.5em}
            \State $\sig[g]{*} \gets \displaystyle \frac{2 - \gamma}{(1 - \gamma) \, v_{g} \, h} + \sig[g]{t}$ \qquad $\hsig[\hg]{*} \gets \displaystyle \frac{2 - \gamma}{(1 - \gamma) \, v_{\hg} \, h} + \hsig[\hg]{t}$
            
            \vspace{0.5em}
            \State $q_{m,g}^{*} \gets q_{m,g}^{(n)} + \displaystyle \frac{1}{\gamma \, (1 - \gamma) \, v_{g} \, h} \, \psi_{m,g}^{(n-1+\gamma)} - \frac{1 -\gamma}{\gamma \, v_{g} \, h} \, \psi_{m,g}^{(n-1)}$
    
            \vspace{0.5em}
            \State $\psi_{m,g}^{(n)} \gets$ Algorithm~\ref{alg:hy-multigroup} with: \Comment{BDF2 Step} 
        
            \hspace{10pt} $\sig[g]{*} \gets \sig[g]{*}$, $\sig[g' \to g]{s} \gets \sig[g' \to g]{s}$, $\chi_{g} \gets \chi_{g}$, $\nu_{g'} \gets \nu_{g'}$, $\sig[g' \to g]{f} \gets \sig[g' \to g]{f}$, $v_g \gets v_{g}$, $q_{m,g}^{*} \gets q_{m,g}^{*}$,

            \hspace{10pt} $\hsig[\hg]{*} \gets \sig[\hg]{*}$, $\hsig[\hg' \to \hg]{s} \gets \hsig[\hg' \to \hg]{s}$, $\hat \chi_{\hg} \gets \hat \chi_{\hg}$, $\hat \nu_{\hg'} \gets \hat \nu_{\hg'}$, $\hsig[\hg' \to \hg]{f} \gets \hsig[\hg' \to \hg]{f}$, $v_{\hg} \gets v_{\hg}$,
            
            \hspace{10pt} $\psi^{(n-1)} \gets \psi^{(n-1+\gamma)}$, $\bsOmega_m \gets \bsOmega_{m}$, $w_m \gets w_{m}$, $\hat{\bsOmega}_{\hm} \gets \hat{\bsOmega}_{\hm}$, $\hw_{\hm} \gets \hw_{\hm}$, 
            
            \hspace{10pt} $\Delta E_g \gets \Delta E_g$, $\Delta \hE_{\hg} \gets \Delta \hE_{\hg}$, $\varepsilon_G \gets \varepsilon_G$, and $\varepsilon_M \gets \varepsilon_M$
        
        \EndFor
        
        \State \Return $\psi_{m,g} \,^{(N)}$
    \end{algorithmic} 
\end{algorithm}

\begin{algorithm}[H]
    \caption{Time step update of the collision-based hybrid multigroup, discrete ordinates equation.}
    \label{alg:hy-multigroup}
    \setstretch{1.4}
    \begin{algorithmic}[1]
    
        \Require $\sig[g]{*}$, $\sig[g' \to g]{s}$, $\chi_{g}$, $\nu_{g'}$, $\sig[g' \to g]{f}$, $v_g$, $q^{*}_{g}$, \Comment{Uncollided Material Properties}
        
        \Require $\hsig[\hg]{*}$, $\hsig[\hg' \to \hg]{s}$, $\hat \chi_{\hg}$, $\hat \nu_{\hg'}$, $\hsig[\hg' \to \hg]{f}$, $v_{\hg}$ \Comment{Collided Material Properties}
        
        \Require $\psi^{(n-1)}$ \Comment{Solution from Previous Time Step}
        
        \Require $\bsOmega_m$, $w_{m}$, $\hat{\bsOmega}_{\hm}$, $\hw_{\hm}$, $\Delta E_g$, $\Delta \hE_{\hg}$ \Comment{Discretization Parameters}
        
        \Require $\varepsilon_G$, $\varepsilon_M$ \Comment{Convergence Tolerances}
        
        \State $\psiu_{m,g} \,^{(n-1)} \gets \psi_{m,g} \,^{(n-1)} \qquad \qu_{g} \gets q^{*}_{g}$
        \For{$g = 1, \dots, G$} \Comment{Uncollided Flux Update}
            \For{$m = 1, \dots, M$} 
                \State $\psiu_{m,g} \gets \left(\bsOmega_m \cdot \nabla + \sig[g]{*}\right)^{-1} \qu_{g}$ \Comment{Transport Sweep}
            \EndFor
        \EndFor
        \State $\barpsiu_g \gets \displaystyle\sum_{m = 1}^{M} w_{m} \psiu_{m,g}$
        
        \vspace{0.5em}
        \State $\qc_{\hg} \gets$
        $\displaystyle\sum_{g = \Gamma_{\hg}+1}^{\Gamma_{\hg+1}} \displaystyle\sum_{g'=1}^{G} \sig[g' \rightarrow g]{s} \barpsiu_{g'} + \displaystyle\sum_{g=\Gamma_{\hg}+1}^{\Gamma_{\hg+1}} \chi_{g} \displaystyle\sum_{g'=1}^{G} \nu_{g'} \sig[g' \rightarrow g]{f} \barpsiu_{g'}$
        \Comment{Collided Source}
        \vspace{0.5em}
        
        \State $\barpsic_{\hm, \hg} \gets$ Algorithm~\ref{alg:multigroup} with: \Comment{Collided Flux Update} 
        
        $\sig[g' \to g]{s} \gets \hsig[\hg' \to \hg]{s}$, $\chi_{g} \gets \hat \chi_{\hg}$, $\nu_{g'} \gets \hat \nu_{\hg'}$, $\sig[g' \to g]{f} \gets \hsig[\hg' \to \hg]{f}$, $v_g \gets v_{\hg}$, $\sig[g]{*} \gets \hsig[\hg]{*}$,
        
        $q^{*}_{g} \gets \qc_{\hg}$, $\psi^{(n-1)} \gets 0$, $\bsOmega_m \gets \hat \bsOmega_{\hm}$, $w_m \gets \hw_{\hm}$, $\varepsilon_G \gets \varepsilon_G$, and $\varepsilon_M \gets \varepsilon_M$
        
        \vspace{0.5em}
        \State $\qt_{m,g} \gets \dfrac{\Delta E_g}{\Delta \hE_{\hg}} \left[\qc_{\hg} +
        \displaystyle\sum_{\hg'=1}^{\hG} \sig[\hg' \rightarrow \hg]{s} \barpsic_{\hg'} + \chi_{\hg} \displaystyle\sum_{\hg'=1}^{\hG} \nu_{\hg'} \sig[\hg' \rightarrow \hg]{f} \barpsic_{\hg'} \right]$
        \Comment{Total Source}
        
        \vspace{0.5em}
        \For{$g = 1, \dots, G$} \Comment{Total Flux Update (Corrector Step)}
            \For{$m = 1, \dots, M$}
                \State $\psit_{m, g} \gets \left(\bsOmega_m \cdot \nabla + \sig[g]{*}\right)^{-1} \qt_{m,g}$ \Comment{Transport Sweep}
            \EndFor
        \EndFor
        \State $\psi_{m, g} \,^{(n)} \gets \psit_{m, g} \qquad \barpsi_g \,^{(n)} \gets \displaystyle\sum_{m = 1}^{M} w_{m} \psit_{m,g}$
        \State \Return $\psi_{m,g} \,^{(n)}$, $\barpsi_{g} \,^{(n)}$
    \end{algorithmic} 
\end{algorithm}

\end{document}